\documentclass[aps,prx,twocolumn,groupedaddress,10pt]{revtex4-2}

\usepackage{amsmath,amsfonts,amssymb}
\usepackage[dvipsnames]{xcolor}
\usepackage{soul}
\usepackage{graphicx}
\usepackage{braket}
\usepackage{cancel} 
\usepackage[normalem]{ulem}

\graphicspath{ {./} }

\newcommand{\EqRef}[1]{Eq.~\eqref{#1}}
\newcommand{\FigRef}[1]{Fig.~\ref{#1}}

\newcommand{\AppRef}[1]{Appendix~\ref{#1}}

\DeclareMathOperator\erf{erf}
\DeclareMathOperator\sgn{sgn}
\DeclareMathOperator\diag{\mathbf{diag}}

\begin{document}

\title{Beyond catastrophic forgetting in associative networks with self-interactions}

\author{Gianni V. Vinci}
\thanks{Equally contributing authors.}
\author{Andrea Galluzzi}
\thanks{Equally contributing authors.}
\author{Maurizio Mattia}
\email{maurizio.mattia@iss.it}

\affiliation{Natl. Center for Radiation Protection and Computational Physics \\ 
Istituto Superiore di Sanità, 00161 Roma, Italy}

\date{\today}

\begin{abstract}
Spin-glass models of associative memories are a cornerstone between statistical physics and theoretical neuroscience.
In these networks, stochastic spin-like units interact through a synaptic matrix shaped by local Hebbian learning.
In absence of self-interactions (i.e., autapses), the free energy reveals catastrophic forgetting of all stored patterns when their number exceeds a critical memory load. 
Here, we bridge the gap with biology by considering networks of deterministic, graded units coupled via the same Amari-Hopfield synaptic matrix, while retaining autapses. 
Contrary to the assumption that self-couplings play a negligible role, we demonstrate that they qualitatively reshape the energy landscape, confining the recurrent dynamics to the subspace hosting the stored patterns. 
This allows for the derivation of an exact overlap-dependent Lyapunov function, valid even for networks with finite size. Moreover, self-interactions generate an auxiliary internal field aligned with the target memory pattern, widening the repertoire of accessible attractor states. 
Consequently, pure recall states act as robust associative memories for any memory load, beyond the critical threshold for catastrophic forgetting observed in spin-glass models---all without requiring nonlocal learning prescriptions or significant reshaping of the Hebbian synaptic matrix.
\end{abstract}


\maketitle

\section{Introduction}

Hebbian learning serves as a fundamental principle in the reorganization of neuronal networks. 
First proposed by D. O. Hebb over seventy years ago \cite{Hebb1949}, it posits that neurons firing together tend to form stronger connections. 
The beauty and the strength of this principle lie in its local nature; the synaptic weight between two neurons adjusts solely based on their individual activities, independent of the firing patterns of other neurons within the network.
Extensive evidence supports the concept of local plasticity in synaptic weights, as documented in neurobiology  \cite{Magee2020}. 
When the activities (i.e., the firing patterns) of pre- and postsynaptic neurons are statistically correlated, synapses are strengthened (potentiated). 
Conversely, when these activities are anticorrelated, synapses weaken (i.e., they are depressed). 
This mechanism underpins the unsupervised learning of the statistical features of the sensed environment.
Indeed, `cell assemblies' activated by similar stimuli tend to fire synchronously, even if only a subset of the neurons is activated  \cite{CarrilloReid2016,Amit1989}. 
Each assembly thus represents a memory linked to all stimuli that elicit the same network state, a concept rooted in the pioneering works of S.-I. Amari and J. J. Hopfield \cite{Amari1972,Hopfield1982}.
According to this theoretical framework also sequential activity patterns can establish statistical correlations that may not be directly relevant to task-solving, as observed in the associative cortices of nonhuman primates \cite{Miyashita1988,Amit1995}. 
This underscores the capacity of Hebbian learning to capture and encode the statistical structure of sensory inputs and inner representations of mental states.

Building on these principles, the statistical physics of recurrent neural networks (RNNs) modeling associative memories, typically considers the Hebbian rule only for connections between couples of  neurons, excluding autapses \cite{Amit1989,Hertz1991}. 
This approach, however, contradicts experimental evidence demonstrating the existence of self-interactions that can lead to potentiated (positive) synapses, consistent with the tenets of Hebbian plasticity \cite{Song2005,Lefort2009}. 
The rationale for excluding self-interactions often stems from the observation that, if these interactions are positive, symmetric synaptic matrices cannot generate cycling activity patterns \cite{Coolen2005}. 
In this case, by setting to zero the self-interactions does not significantly impair the performances of RNNs as associative memories \cite{Kanter1987,Hertz1991}.
This belief is underpinned by the expectation that in large RNNs, the contribution from self-interactions becomes negligible, scaling with the number of units $N$ in the network as $\sim 1/\sqrt{N}$ \cite{Kanter1987,Clark2024}.
However, RNNs with strong self-interactions can exhibit attractor dynamics related to encoded memories, even when the synaptic matrix is asymmetric \cite{Coolen2005,Stern2014}. 
This suggests that self-interactions may play a more critical role in the dynamics of RNNs than previously acknowledged \cite{Stern2023}.

Here, we investigate this role in RNNs modeling associative memories incorporating the Amari-Hopfield synaptic matrix \cite{Amari1972,Hopfield1982} with self-interactions (self-couplings). 
Our findings indicate that self-couplings play a crucial role in stabilizing existing fixed points by expanding their basins of attraction. 
This enhancement contributes to the network's robustness in recalling stored patterns (i.e., memories), a critical characteristic for effective memory storage and retrieval systems.
Moreover, this stability addresses the long-standing `blackout catastrophe' problem \cite{Amit1989}.
When the number $P$ of independent patterns stored in the synaptic matrix exceeds a critical threshold of $\sim 0.14 N$ \cite{Amit1985,Crisanti1986}, memories can no longer be recalled, leading to what is known as `catastrophic forgetting.'
Our observations challenge traditional perspectives by suggesting that neuronal networks with self-interactions can achieve maximal memory capacities without relying on nonlocal synapses \cite{Personnaz1985,Kanter1987} or novel learning rules \cite{Hopfield1983,Diederich1987}.

\section{Hebbian RNNs with self-couplings}

Our reference model of associative memory is a RNN with $N$ units each with activity $r_i$ evolving deterministically in time as
\begin{equation}
   \tau \dot{r}_i = \Phi(h_i) - r_i
\label{eq:SingleUnitDyn}
\end{equation}
for any $i \in [1,N] \subset \mathbb{Z}$.
Unit activities represent the instantaneous number of spikes emitted per unit time (i.e., the firing rate) by a local and homogenous assembly of neurons.
Indeed, under mean-field approximation, each of these assemblies composed of excitatory and inhibitory neurons, can be effectively described by a similar low-dimensional dynamics \cite{Mattia2002,Abbott1993,Treves1993,Brunel1999,Knight2000}.
The activity of each unit is driven by a current-to-rate gain $\Phi(h_i)$, a monotonically increasing function ($\Phi' \geq 0$) amplifying the synaptic input $h_i$.
In absence of any driving force ($\Phi = 0$), firing rates relax to 0 with decay time $\tau$.
Synaptic input is determined by a linear combination of the `presynaptic' unit activities weighted by the synaptic matrix $\mathbf{J} \in \mathbb{R}^{N \times N}$:
\begin{equation}
   h_i = \sum_{j=1}^N{J_{ij} r_j} + h^{(\mathrm{ext})}_i \, .
\label{eq:SynapticInput}
\end{equation} 
For the sake of generality, here we included also the contribution of an external input $h^{(\mathrm{ext})}_i$, although we will only deal with isolated RNNs ($h^{(\mathrm{ext})}_i = 0$).

The introduced network dynamics \eqref{eq:SingleUnitDyn} and \eqref{eq:SynapticInput} can be conveniently rewritten in matrix formalism as 
\begin{equation}
\begin{split}
   \tau \ket{\dot{r}} & = \ket{\Phi(h)} - \ket{r} \\
   \ket{h} & = \mathbf{J}\ket{r} + \ket{h^{(\mathrm{ext})}} \, , 
\end{split}
\label{eq:RNNdyn}
\end{equation}
by resorting to the `bra-ket' notation used in quantum mechanics where, for instance, the `ket' $\ket{\Phi(h)} \in \mathbb{R}^{N \times 1}$ is a column vector with elements $\Phi(h_i)$.
Note that, under the hypothesis of a stationary external input ($\ket{\dot{h}^{(\mathrm{ext})}} = 0$), a self-consistent dynamics for the synaptic currents results \cite{Shiino1993}:
\begin{equation}
   \tau \ket{\dot{h}}= \mathbf{J}\ket{\Phi(h)} - \ket{h} + \ket{h^{(\mathrm{ext})}} \, ,
\label{eq:hDynamics}
\end{equation}
which in turn is equivalent to the one of the well-known associative networks of graded (analog) `neurons' introduced in \cite{Cohen1983,Hopfield1984,Kuhn1991}.

Under specific conditions, these networks can store $P$ patterns $\ket{\xi_\mu}$ of activity ($\mu \in [1,P] \subset \mathbb{Z}$) as stable equilibrium points of the collective dynamics by properly designing their synaptic matrix \cite{Amari1972,Hopfield1982}.
Following a Hebbian prescription for synaptic plasticity, synaptic couplings $J_{ij}$ are set to be proportional to the correlation of pre- and postsynaptic activity ($r_i$ and $r_j$, respectively) of presented patterns
\begin{equation}
   J_{ij} = \frac{g}{N} \sum_{\mu=1}^P{\xi_{\mu i}\xi_{\mu j}} \, .
\label{eq:AHWij}
\end{equation}
Assuming the patterns to memorize as independent vectors, the synaptic matrix is a sum of $P$ rank-1 matrices
\begin{equation}
   \mathbf{J} = g \sum_{\mu=1}^P{\ket{\xi_\mu}\bra{\xi_\mu}} \, ,
\label{eq:AHSynMat}
\end{equation}
where the `bra' $\bra{\xi_\mu} = \ket{\xi_\mu}^\intercal/N$ is a row vectors proportional to the transpose of the $\mu$-th pattern.
Here, $g$ is the relative strength of synaptic couplings, and plays the role of the inverse of a temperature in spin-glass models of neural networks \cite{Kuhn1991}. 
According to \cite{Amari1972,Hopfield1982}, patterns have random elements $\xi_{\mu i} \in \{-1,+1\}$, such that in the thermodynamic limit $N \to \infty$, they compose an orthonormal basis of a $P$-dimensional space.
Indeed, in infinite-size networks $\braket{\xi_\mu|\xi_\nu} \equiv \sum_{i=1}^N \xi_{\mu i}\xi_{\nu i}/N = \mathbb{E}[\braket{\xi_\mu|\xi_\nu}] \underset{N\to\infty}{\rightarrow} \delta_{\mu\nu}$ being the variance $\mathbb{V}[\braket{\xi_\mu|\xi_\nu}] = (1-\delta_{\mu\nu})/N$ for any $\mu, \nu \in [1,P]$.
Note that with this choice of patterns, the self-couplings (i.e., the diagonal elements of $\mathbf{J}$) are
\begin{equation}
  J_{ii} = g \, \alpha \, ,
\label{eq:SelfCouplings}
\end{equation}
i.e., positive and proportional to the `memory load' $\alpha \equiv P/N$. 
This differs from the usual setting $J_{ii} = 0$ adopted in statistical mechanics \cite{Amit1989}.
In the following Sections we will show that such difference has not a negligible impact. 

As these RNNs aim at modeling biological networks of local assemblies of neurons, the less `realistic' element we are going to incorporate is the current-to-rate function (i.e., the `graded-response'), which according to previous works is 
\begin{equation}
   \Phi(h_i) \equiv \tanh(h_i)  \, .
\label{eq:Tanh}
\end{equation}
Indeed, with this choice the equilibrium points in \EqRef{eq:RNNdyn} have $r_i \in [-1,+1]$. 
However, firing rates cannot be negative and for this reason, $r_i$ should be rather seen as a rescaled frequency of emitted spikes \cite{Amit1989}.

\section{Low-dimensional $\vec{m}$-dynamics}

The main impact of having a $\mathbf{J}$ with non-null diagonal elements as in \EqRef{eq:SelfCouplings}, is to confine the RNN dynamics into a low-dimensional space.
To prove it we introduce the `overlap' $m_\mu$ (i.e., the projection) of the network activity with the $\mu$-th pattern
\begin{equation}
   m_\mu = \braket{\xi_\mu|r} = \frac{1}{N}\sum_{i=1}^N \xi_{\mu i} r_i \, .
\label{eq:Overlaps}
\end{equation} 
The synaptic input in \EqRef{eq:SynapticInput} of an isolated network ($h^{(\mathrm{ext})}_i = 0$) result to be a function of these overlaps:
\begin{equation}
\begin{split}
	\ket{h} =& \, \mathbf{J}\ket{r} = g \sum_{\mu=1}^P \ket{\xi_\mu}\braket{\xi_\mu|r} \\
           =& \, g \sum_{\mu=1}^P m_\mu \ket{\xi_\mu} \equiv g \ket{\mathbf{\Xi} \vec{m}} \, .
\end{split}
\label{eq:MSynInput}
\end{equation}
Here $\mathbf{\Xi} \in \mathbb{R}^{N \times P}$ is the rectangular matrix whose columns are the $P$ patterns $\ket{\xi_\mu}$, and $\vec{m} \in \mathbb{R}^{P \times 1}$ is the $P$-dimensional column vector of the overlaps $m_\mu$.
Now applying to both hand sides of the $r$-dynamics \eqref{eq:RNNdyn} the `bra' $\bra{\xi_\mu}$, we eventually obtain a self-consistent dynamics for the overlaps:
\begin{equation}
   \tau \dot{m}_\mu = \braket{\xi_\mu|\Phi(\vec{m})} - m_\mu \, ,
\label{eq:SingleMDyn}
\end{equation} 
where $\ket{\Phi(h)} = \ket{\Phi(\vec{m})}$ being $h_i = h_i(\vec{m}) = g \{\ket{\mathbf{\Xi} \vec{m}}\}_i$.
In matrix form this equation reduces to
\begin{equation}
   \tau \dot{\vec{m}} = \mathbf{\Xi}^+ \ket{\Phi(\vec{m})} - \vec{m} \, ,
\label{eq:MDyn}
\end{equation} 
where $\mathbf{\Xi}^+ = \mathbf{\Xi}^\intercal/N \in \mathbb{R}^{P \times N}$, which in the limit $N\to\infty$ is the pseudoinverse of $\mathbf{\Xi}$ as $\mathbf{\Xi}^+ \mathbf{\Xi}=\mathbf{I}_P$ (i.e., the identity matrix $P \times P$), leading to satisfy the Moore–Penrose conditions.

Of course the vector $\ket{\Phi(\vec{m})}$ is still living into the $N$-dimensional space of the full RNN dynamics.
As a consequence, $\ket{\dot{r}}$ is expected to drive the whole network activity outside the $P$-dimensional space defined by the patterns. 
In order to describe this component, we decompose the network activity as combination of two contributions:
\begin{equation}
  \ket{r} \equiv \ket{\mathbf{\tilde{\Xi}}\vec{m}} + \ket{r_\perp} \, ,
\label{eq:OrthogActivity}
\end{equation}
where the residual activity $\ket{r_\perp}$ is orthogonal to the pattern subspace: $\braket{\xi_\mu|r_\perp} = 0$ for any $\mu \in [1,P]$.
Here, $\mathbf{\tilde{\Xi}} \equiv N \mathbf{Q} \mathbf{R}^\intercal$ with $\mathbf{Q} = \mathbf{\Xi} \mathbf{R}$. 
$\mathbf{Q}$ is the matrix composed by the versors forming the basis of the pattern subspace.
These $P$ versors result from the Gram-Schmidt orthonormalization of the $\ket{\xi_\mu}$ through a well-defined linear transform $\mathbf{R} \in \mathbb{R}^{P \times P}$.
Note that this is true for any network size $N$, and the projector to the pattern subspace can be explicitly written as $\mathbf{Q} \mathbf{Q}^\intercal = \mathbf{\tilde{\Xi}} \mathbf{\Xi}^+$ such that $\ket{r_\perp} = (\mathbf{I}_N - \mathbf{\tilde{\Xi}} \mathbf{\Xi}^+) \ket{r}$. 
 
By deriving with respect to time both hand sides of \EqRef{eq:OrthogActivity}, and making use of Eqs.~\eqref{eq:RNNdyn} and \eqref{eq:MDyn}, the dynamics of the residual activity is
\begin{equation}
   \tau \ket{\dot{r}_\perp} = (\mathbf{I}_N - \mathbf{\tilde{\Xi}} \mathbf{\Xi}^+) \ket{\Phi(\vec{m})} - \ket{r_\perp} \, .
\label{eq:ResActivityDyn}
\end{equation}
This equation together with \EqRef{eq:MDyn} are equivalent to \EqRef{eq:RNNdyn}, and are exact for any network size $N$.
However, it is apparent from these new expressions the existence of two subsystems working in a `master-slave' regime, where the $\vec{m}$ state evolves independently from the residual activity $\ket{r_\perp}$, while the opposite is not true.
As a consequence, the dynamics in the $(N-P)$-dimensional subspace hosting $\ket{r_\perp}$ has no role in determining the fixed-points and the related activity relaxation of the RNN, as it is fully described by \EqRef{eq:MDyn}. 
\begin{figure}
   \centering
   \includegraphics[width=86mm]{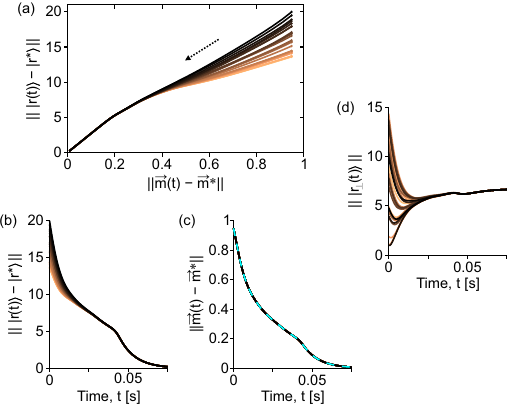}
   \caption{Numerical integration of \EqRef{eq:RNNdyn} with the Amari-Hopfield synaptic matrix \eqref{eq:AHSynMat} from 25 independent initial states $\ket{r(0)}$ with fixed $\ket{\tilde{\mathbf{\Xi}}\vec{m}(0)}$ and randomly sampled $\ket{r_\perp(0)}$.
   (a) Neural trajectories in the plane of the distances between the network state and the ending equilibrium point ($||\ket{r(t)} - \ket{r^*}||$) and the distances of their projections on the pattern subspace ($\vec{m}(t) = \mathbf{\Xi}^+\ket{r(t)}$).
   (b-c) Same distances versus time.
   Dashed line in (c), $\vec{m}$-trajectory from \EqRef{eq:MDyn} with same initial condition $\vec{m}(0)$.
   (d) Euclidean norm of the orthogonal component $\ket{r_\perp(t)}$ from the same neural trajectories. 
	Relevant parameters: $N = 200$, $P = 10$ (i.e., $\alpha = 0.05$), $\tau = 0.010$ and $g = 20$. 
	}
   \label{fig:RelaxDynPdim}
\end{figure}
This scenario is well illustrated in \FigRef{fig:RelaxDynPdim}(a)-(c), where the dynamics of an RNN starting from different initial states $\ket{r(0)}$ relaxes always to the same equilibrium point $\ket{r^*} = \ket{\Phi(r^*)}$.
In this example case the network state at $t=0$ is chosen by randomly sampling $\ket{r_\perp(0)}$ and setting always the same $\vec{m}(0)$.
As expected, $\vec{m}(t)$ displays always the same trajectory in approaching $\vec{m}^* = \mathbf{\Xi}^+\ket{r^*}$ [\FigRef{fig:RelaxDynPdim}(c)], differently from $\ket{r(t)}$ and $\ket{r_\perp(t)}$ [\FigRef{fig:RelaxDynPdim}(b) and (d), respectively].
Not surprisingly, integrating \EqRef{eq:MDyn} with same $\vec{m}(0)$ leads exaclty to the same relaxation dynamics in the pattern subspace [dashed line in \FigRef{fig:RelaxDynPdim}(c)].

Here, it is important to remark that such master-slave dynamics does not emerge in the case $J_{ii} = 0$.
Indeed, if the synaptic matrix has null diagonal elements as usually assumed in spin-glass models, $\mathbf{J} \to \mathbf{J} - g \, \alpha \mathbf{I}_N$ and \EqRef{eq:MDyn} will include an explicit dependence on $\ket{r_\perp}$ making the two subsystems reciprocally coupled.

\section{$\vec{m}$-based Lyapunov function}

The introduced RNN is a deterministic system dissipating the stored energy, and being $\mathbf{J}$ symmetric, it always approaches a stable equilibrium point.
For these systems, Lyapunov functions $E$ represent such stored energy. 
They always decrease in time ($\dot{E}\leq0$), unless the network state is a fixed-point where $E$ is constant. 
As the dynamics of these RNNs is fully determined by the overlaps $\vec{m}$, the Lyapunov function result to be (see the \AppRef{app:LyapFunc})
\begin{equation}
   E(\vec{m}) = \frac{1}{2} \sum_{\mu = 1}^P m_\mu^2 - \frac{1}{g N} \sum_{i=1}^N F(h_i) \, ,
\label{eq:LyapunovFunc}
\end{equation}
where $F(h) = \int \Phi(h) dh$ and no external input is taken into account ($h_i^{(\mathrm{ext})} = 0$).
For the specific graded-response in \EqRef{eq:Tanh} the primitive $F$ can be explicitly carried out giving
\begin{equation}
E(\vec{m}) = \frac{1}{2} \sum_{\mu = 1}^P m_\mu^2 - \frac{1}{g N} \sum_{i=1}^N \log(\cosh(h_i)) \, ,
\label{eq:LyapunovLogCosh}
\end{equation}
with synaptic input $h_i(\vec{m})$ from \EqRef{eq:MSynInput}.
This result is exact and does not require any thermodynamic limit ($N \to \infty$).
This Lyapunov function in the $\vec{m}$-space is complemented by the well-known expression for the stored energy derived in \cite{Cohen1983,Hopfield1984} and defined in the $h$-space of the synaptic currents:
\begin{equation}
   E(\ket{h}) = \frac{1}{2} \sum_{i,j = 1}^N J_{ij} \Phi(h_i) \Phi(h_j) + \sum_{i,j = 1}^N \left[ h_i \Phi(h_i) - F(h_i) \right]
\label{eq:LyapunovCGH}
\end{equation}
where for the mentioned graded-response \eqref{eq:Tanh} we must consider again $F(h_i) = \log(\cosh(h_i))$.

\begin{figure}[htp]
    \centering
    \includegraphics[width=86mm]{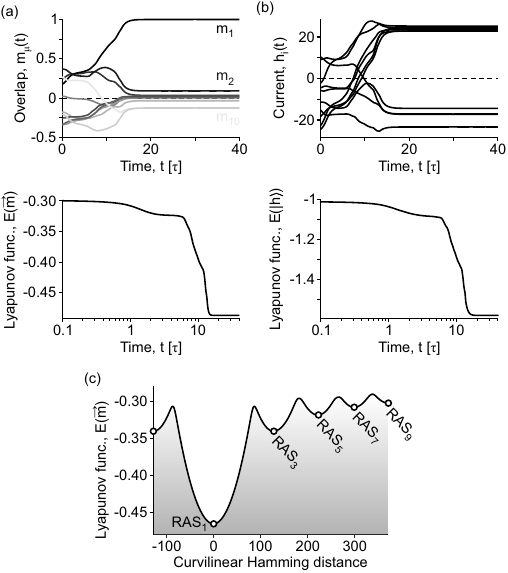}
    \caption{Lyapunov functions of a RNN relaxing into the attractor state `recalling' the pattern $\ket{\xi_1}$ (RAS$_1$).
   (a) Relaxation dynamics (top) of the $P=10$ overlaps $m_\mu$ of a RNN composed of $N=500$ units, with synaptic matrix \eqref{eq:AHSynMat} and relative strength $g=20$.
   Bottom, Lyapunov function $E(\vec{m})$ in the $m$-space from \EqRef{eq:LyapunovLogCosh} monotonically decaying as the equilibrium point is approached.
   (b) Same as (a) but showing the first 10 $h_i(t)$ (top) and current-based Lyapunov function $E(\ket{h})$ from \EqRef{eq:LyapunovCGH}.
   (c) Overlap-based $E(\vec{m})$ computed along the polygonal chain connecting the attractor states of pure recall (RAS$_1$) and the symmetric mixtures RAS$_n$ reached starting from $\ket{r(0)}=\sgn(\sum_{\mu=1}^n \ket{\xi_\mu})$.}
    \label{fig:LyapunovFunc}
\end{figure}

For an example RNN, the always decaying $E(\vec{m})$ and $E(\ket{h})$
are shown in \FigRef{fig:LyapunovFunc}(a) and (b), respectively.
As $\alpha = 0.02 \ll 1$ (low memory loading \cite{Amit1989,Kuhn1991}), the network state can relaxes into the so-called `recall' attractor state (RAS), where single-unit activities aligns with pattern $\ket{\xi_1}$ in this case, such that only the overlap $m_1$ is close to 1 [\FigRef{fig:LyapunovFunc}(a)-top] with synaptic currents polarized to relatively large (positive and negative) values [\FigRef{fig:LyapunovFunc}(b)-top].
In spin-glass models for small enough $\alpha$, besides the pure recall state, other equilibrium points are accessible named symmetric mixture states.
In these spurious attractor states -- called here RAS$_n$ -- large and equal overlaps with an odd number $n$ of patterns emerge, while the remaining $m_\mu$ fluctuate around 0 \cite{Amit1989}.
We found the same richness of spurious states in the example RNN [\FigRef{fig:LyapunovFunc}(c)], recovering the expected hierarchy of energies found in spin-glass models \cite{Amit1989} where RAS$_1$ (pure recall) displays the lowest $E(\vec{m})$, and with higher and higher energies for RAS$_n$ with increasing size $n$ of the mixtures.

As $E(\ket{h})$ describes the relaxation dynamics of the full network state, including the orthogonal component $\ket{r_\perp(t)}$, it is not surprising that $E(\vec{m})$ is a different function. 
However, for what found in the previous Section, the current-based Lyapunov function does not add any information that cannot be recovered from what found in the $\vec{m}$-space, eventually leading $E(\vec{m})$ to be a complete representation of the RNN dynamics \eqref{eq:MDyn}.

It is interesting to note that \EqRef{eq:LyapunovLogCosh} is remarkably similar to the free energy obtained in the thermodynamic limit for attractor networks of spin-like stochastic neurons with a finite number $P$ of stored patterns (i.e., $\alpha \to 0$, $N \to \infty$ and $J_{ii} = 0$). 
Indeed, under mean-field approximation the free energy of these models is equal to our $E(\vec{m})$ with the only difference of a factor of 2 multiplying $\cosh(h_i)$ \cite{Amit1985b,Amit1989}. 
It has been also proven that this free energy can be evaluated in terms of eigenvalues and eigenvectors of the synaptic matrix leading to a limited number of order parameters \cite{VanHemmen1986,Grensing1987}.
In the same limit, but relying on a master-equation approach for the probability density of the network states, the same Lyapunov function in \EqRef{eq:LyapunovLogCosh} was derived \cite{Coolen2005}.
Such stochastic approach assuming the asynchronous updates of the spins did not require any constrains on $P$, and self-couplings was as in \EqRef{eq:SelfCouplings} eventually leading to the same dynamics \eqref{eq:MDyn} we derived for the overlaps.
We believe this is a strong indication that in the $N \to \infty$ limit, the role of self-couplings in the RNNs we are going to present for any $N$ does not depend on the fact that the single-unit dynamics is continuous and deterministic or discrete and stochastic.

\section{Recall states at large-$g$}

The equilibrium points where the RNN dynamics relaxes, i.e. the minima of the Lyapunov function \eqref{eq:LyapunovFunc}, are those with $\dot{\vec{m}} = 0$ in \EqRef{eq:MDyn}, i.e.,
\begin{equation}
   \vec{m} =\mathbf{\Xi}^+ \ket{\Phi(g \mathbf{\Xi} \vec{m})} \, .
   \label{eq:FixedPoints}
\end{equation}
This is the same mean-field ($N \to \infty$) equation for self-consistency of the overlaps at equilibrium in spin-glass models at low memory load ($\alpha \to 0$) \cite{Amit1985b,Amit1989}. 
In this limit the two models are then equivalent allowing to interpret the relative strength $g$ as the inverse $1/T$ of the temperature \cite{Kuhn1991}. 
Thus referring to the known results from the statistical mechanics at $T=0$, here we focus on the recall capabilities in the limit $g \to \infty$.

In this limit a recall state RAS$_1$ can be deterministically approached by setting as initial condition one of the stored patterns. 
Without loss of generality, we choose the first pattern as initial condition $\ket{r(0)} = \ket{\xi_1}$, eventually relaxing to a recall state with $m_1 \gg m_{\mu>1}$.
In the $N\to\infty$ limit, the overlaps $m_{\mu>1}$ of the non-recalled (i.e., `uncondensed') patterns have Gaussian distribution with $0$ mean and variance $v/N$ with 
\begin{equation}
   v = \frac{1}{\alpha} \sum_{\mu=2}^P m_\mu^2 \, .
   \label{eq:UncondNoiseVar}
\end{equation}
This is reminiscent of the parameter `$r$' describing in spin-glass models the noise due to the uncondensed patterns \cite{Amit1989}.

\begin{figure}
   \centering
   \includegraphics[width=86mm]{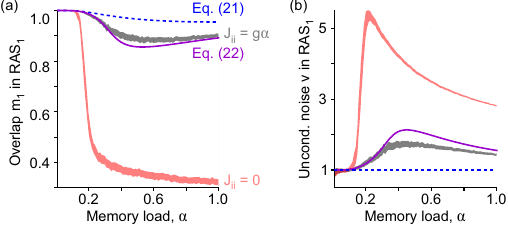}
   \caption{Memory recall at $g \to \infty$. 
   (a) Overlap $m_1$ (mean $\pm 5$ SEM) at equilibrium over 300 different realizations RNNs with $N = 500$ by random sampling the $P = \alpha N$ patterns.
   Initial condition, $\ket{r(0)} = \ket{\xi_1}$. 
   Gray and red, overlap $m_1(\alpha)$ in the RAS$_1$ with intact synaptic matrix ($J_{ii}=g \alpha$) and the one with $J_{ii} = 0$ as usually set in spin-glass models.
   Dashed-blue, overlap from approximated solution in \EqRef{eq:m1LowAlpha}.
   Solid-purple, overlap $m_1$ from replica theory in \EqRef{eq:PredictionMR}.
   (b) Noise size of the uncondensed patterns $v(\alpha)$.
   Colors and line style as in (a).
   As $g \to \infty$, in simulations we changed $\tanh(g \, x) \to \sgn(x)$. 
   Network dynamics \eqref{eq:RNNdyn} is integrated numerically resorting tothe Euler method with a step $dt = 10^{-5} \tau$. 
   Total integration time, $2000 \tau$.}   
	\label{fig:RecallT0}
\end{figure}

A rough approximation to the solution of the self-consistency \EqRef{eq:FixedPoints} can be obtained by applying the Newton–Raphson method.
As a first step, the velocity field in \EqRef{eq:MDyn} computed on the starting state with $m_1^{(0)} = 1$ and $m_{\mu>1}^{(0)} = \braket{\xi_\mu|\xi_1}$, provides a first refined estimate of the searched equilibrium point
\begin{equation}
   m_\mu^{(1)} = \braket{\xi_\mu|\sgn(\mathbf{\Xi} \vec{m}^{(0)})} \, .
   \label{eq:IterativeM}
\end{equation}
The first step has an exact analytical expression in the thermodynamic limit $N\to\infty$ under the hypothesis of uncorrelated patterns (i.e., $\alpha \to 0$, see \AppRef{app:LowAlphaSCE} for details).
Indeed, in this case the uncondensed noise $v^{(0)} = 1$ and the first-step overlap is
\begin{equation}
    m_1^{(1)} = \erf\left( \frac{1+\alpha}{\sqrt{2 v^{(0)} \alpha}} 
 \right) = \erf\left(\frac{1+\alpha}{\sqrt{2 \alpha}}\right) \, ,
   \label{eq:m1LowAlpha}
\end{equation}
similarly to what found in spin-glass models at zero temperature \cite{Krauth1988,Folli2017}. 
Although this approximation may fail for not vanishing memory loads, it is interesting to note that it is a non-monotic function displaying a minimum at $\alpha=1$ with $m_1^{(0)} = \erf(\sqrt{2}) \simeq 0.95$ [\FigRef{fig:RecallT0}(a), dashed line].

This result might suggest the possibility to have memory recall beyond the known capacity limit $\alpha_c \approx 0.14$ in spin-glass models at $T=0$ \cite{Amit1985,Amit1989}.
To confirm such hypothesis, we must take into account pattern correlations resorting to traditional results of replica theory \cite{Amit1989}.  
To facilitate this task, we note that the non-null diagonal elements of $\mathbf{J}$ acts as an external field $\ket{h^\mathrm{(ext)}} = g \alpha \ket{\xi_1}$ if we set $J_{ii} = 0$. 
Self-consistency mean-field equations for this specific case are then \cite{Amit1989}
\begin{equation*}
\begin{split}
   m_1 &=\int_{-\infty}^{\infty} \tanh(g m_1 z) d\mu_{\alpha,v}(z) \\
   q &= \int_{-\infty}^{\infty} \tanh(g m_1 z)^2 d\mu_{\alpha,v}(z) \\
   v &=\frac{q}{(1-g(1-q))^2}
\end{split} \, ,
\end{equation*}
where $d\mu_{\alpha,v} = \exp\left[\frac{(z-\alpha -1)^2}{2 v \alpha}\right] dz/\sqrt{2 \pi v \alpha}$ and $q$ is the Edwards-Anderson order parameter \cite{Edwards1975}. 
In the limit $g \to \infty$ the above system simplifies in
\begin{equation}
\begin{split}
   m_1 &= \erf \left (\frac{\alpha+1}{\sqrt{2 v \alpha}} \right) \\
   v   &= \left[ 1-\frac{1}{m_1} \sqrt{\frac{2}{\pi v \alpha}} e^{\frac{(\alpha +1)^2}{2 v \alpha}} \right]^{-2}
\end{split} \, .
\label{eq:PredictionMR}
\end{equation}

\begin{figure*}[!htb]
   \centering
   \includegraphics[width=179mm]{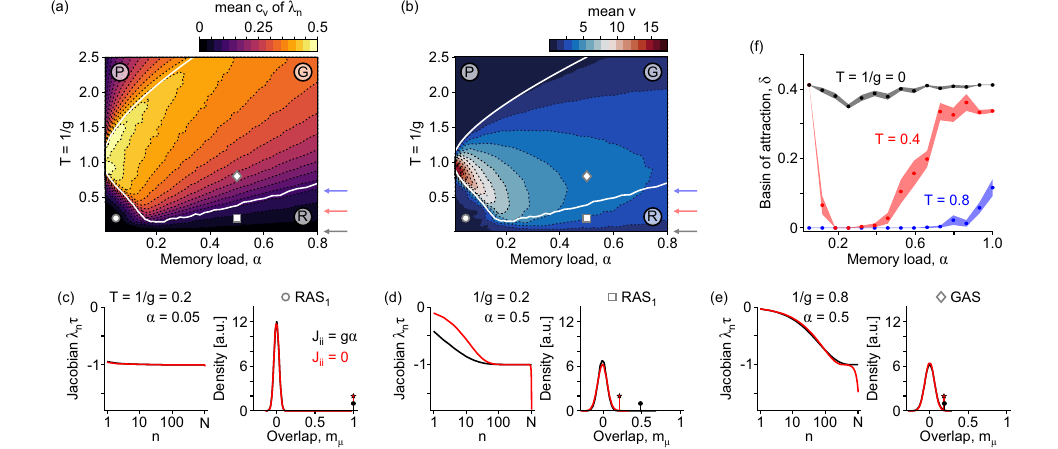}
   \caption{Phase diagram $(\alpha,1/g)$.
   Mean coefficient of variation ($c_v$) of the Jacobian  eigenvalues (a) and size $v$ of the uncondensed noise (b) in the equilibrium point where the RNNs relax after a long transient $\sim 10^3 \tau$.
   Numerical integrations of the RNN dynamics starts with $\ket{r(0)} = \ket{\xi_1}$. 
   Averages performed for each $g$ and $\alpha$ over 100 different realizations of the synaptic matrix $\mathbb{J}$.
   White boundaries separate paramagnetic, glassy and recall phases (from top to bottom, respectively).
	(c-e) Eigenvalues of the Jacobian matrix (left) and distribution of overlap $m_\mu$ (right) at the equilibrium point $\vec{m}^*$ for two recall state RAS$_1$ (circle and square in (a) and (b), $1/g = 0.2$ and $\alpha = 0.05$ and $0.5$, respectively), a glassy state (GAS, diamond, $1/g = 0.8$ and $\alpha = 0.5$).
   Red, eigenvalues and distributions for the same RNNs without self-couplings ($J_{ii} = 0$).
   Dots and stars, maximum values of $m_1$ with or without self-couplings. 
	(f) Size $\delta$ of the basin of attraction of the recall state RAS$_1$, averaged over 100 realizations (dots). 
   Shaded strips, 3 SEM for $T = 0$, $0.4$ and $0.8$. 
	Network size, $N=1000$.}
    \label{fig:Diagram}
\end{figure*}

This novel expression demonstrates strong agreement with numerical simulations of finite-size RNNs, accurately capturing both the overlap $m_1$ and the variance $v$ of the uncondensed noise [\FigRef{fig:RecallT0}(a) and (b), respectively].
For comparison, we include simulation results for the common RNNs without self-couplings ($J_{ii} = 0$), which reproduce established phenomena such as the abrupt decline in $m_1$ near critical memory loads ($\alpha_c \simeq 0.14$).
The inclusion of Hebbian self-couplings ($J_{ii} = g \alpha$) fundamentally alters this behavior mitigating the catastrophic forgetting, with an almost perfect recall ($m_1 \simeq 1$) even at high memory loads.
As shown in \FigRef{fig:RecallT0}(b), this is due to a significant attenuation of the uncondensed noise determined by pattern correlations.
Self-couplings generate an auxiliary internal field aligned with the target memory pattern, likely stabilizing the related RAS$_1$.

\section{Phase diagram $(\alpha,1/g)$}

Consider now the same RNN but with finite $g$, a condition that is more realistic for network models of associative memory.
Analogously to spin-glass models with graded neurons \cite{Kuhn1991}, we expect to find a `paramagnetic' phase with only a single attractor state (PAS) with null overlaps: $m_\mu = 0$.
In the phase diagram $(\alpha,1/g)$ this state can be found where the dynamics \eqref{eq:RNNdyn}, linearized in the vicinity of the related equilibrium point $\ket{m^*}=\ket{0}$, is stable. 
This happens when the maximum eigenvalue $\lambda_{\max}$ of the Jacobian matrix $[\diag(\ket{\Phi'(\vec{m}^*)}) \, \mathbf{J} - \mathbf{I}_N]/\tau$ is negative.
As $\Phi'(0) = 1$, in the thermodynamic limit ($N\to\infty$) 
\begin{equation*}
   \tau \lambda_{\max} = g \lambda_{\max}(\mathbf{\Xi \Xi^{+}})-1 = g(1+\sqrt{\alpha})^2 - 1 \, .
\end{equation*}
Here $\mathbf{J} = g \mathbf{\Xi \Xi^{+}}$, and a standard result from random matrix theory has been used \cite{Akemann2015}. 
Thus PAS loses stability at
\begin{equation}
   \frac{1}{g_c} = (1+\sqrt{\alpha})^2 \, .
   \label{eq:ParamagPhase}
\end{equation}
The inclusion of self-couplings then enhances the critical `temperature' $T_c = 1/g_c$ by a factor of $1 + \sqrt{\alpha}$ relative to standard spin-glass models \cite{Kuhn1991,Amit1995}, thereby reducing the paramagnetic phase domain in the $(\alpha,1/g)$ phase diagram [top boundary in \FigRef{fig:Diagram}(a-b)].

To establish numerically the boundary limiting the recall phase where the RAS$_1$ exists, we ask that $m_1$ at equilibrium reached from $\ket{\xi_1}$ is statistically distinguishable from the overlaps with all other patterns. 
In the large-$N$ limit, the distribution of $m_{\mu>1}$ in RAS$_1$ is Gaussian with 0 mean and variance $v/N$. 
We then consider the pattern $\ket{\xi_1}$ as recalled if $|m_1| > 5 \sqrt{v/N}$ [bottom boundary in \FigRef{fig:Diagram}(a-b)].
With this we find that the recall phase extends for any $\alpha$ also to the case of large but finite synaptic strengths ($g>5$). 
Not only, for $\alpha>0.2$ the recall phase increases linearly its domain with the memory load, i.e.,  
\begin{equation}
   \frac{1}{g_c} \simeq \alpha \, .
   \label{eq:RecallPhase}
\end{equation}
Self-couplings have then a stabilizing role for memory recall. 
This is in part due to the induction of an effective additional field $\sim g \alpha \ket{\xi_1}$.
This recurrent input stabilizes RAS$_1$ provided that it is strong enough: $g \, \alpha \gtrsim 1$.
This endogenous field is associated to the `flip-flop' behavior of the units in the network with self-couplings $J_{ii} > 1$ \cite{Stern2014}.
Indeed, under this condition, $r_i(t)$ evolves with dissipative dynamics equivalent to those of a particle in a double-well energy landscape shaped by the instantaneous current $h_i(t)$, excluding self-interaction.
Given an initial condition close to the pattern to recall, the network dynamics is then biased by the aforementioned self-coupling field, which in turn exhibits resilience to perturbations due to the energy barrier separating positive and negative $r_i(t)$ states.

According to spin-glass models \cite{Amit1989,Kuhn1991}, between paramagnetic and recall phase we found a `glassy' region, where $v \gg 1$ and $m_\mu \ll 1$ for all patterns, at least for not too large $T$ [\FigRef{fig:Diagram}(b)].
In this phase, the network relax into a glassy attractor state (GAS) where no overlap is statistically distinguishable from the others [\FigRef{fig:Diagram}(e)-right]. 
The eigenvalues of the Jacobian in this state are maximally diverse [compare \FigRef{fig:Diagram}(e) with (c) and (d), for a RAS$_1$ and PAS, respectively].
As the Hessian of the corresponding minimum of $E(\vec{m}^*)$ has same eigenvalues of the Jacobian but with opposite sign, the related energy landscape is rather `rough.'

The degree of roughness is well-captured by the coefficient of variation $c_v = \sqrt{\mathbb{V}[\lambda_n]}/\mathbb{E}[\lambda_n]$, and it differently represents the nature of PAS, GAS and RAS$_1$ [\FigRef{fig:Diagram}(a)].
Indeed, memorized patterns are associated with narrow minima that exhibit a nearly perfect paraboloid shape [$\lambda_n \simeq -1/\tau$, \FigRef{fig:Diagram}(c,d)].
Differently, PAS is in a minimum centered at the origin of the $\vec{m}$-space with increasing roughness as the glassy phase is approached, as $c_v^{(\mathrm{PAS})} = \sqrt{\alpha}/(T-1)$ with $T = 1/g$. 
From this perspective, the role of self-couplings is even more apparent by comparing the spectrum of our Jacobian (black) and the one computed imposing $J_{ii} = 0$ ($\vec{m}^*$ is the same) in the new recall region at large $\alpha$ [\FigRef{fig:Diagram}(d)].
The absence of self-couplings facilitates the interference with nearby minima, widening the distribution of $\lambda_n$, and flattening the curvature in specific directions. 
So far in studying the stability of the RAS$_1$ we used as initial state the pattern to recall. 
However, associative memories compress sets of input into a single prototype. 
The newly found RAS$_1$ must then have a relatively wide basin of attraction.
As in classical studies of spin-glass models, we define the basin of attraction of a RAS$_1$ as the maximum fraction $\delta$ of `spin' (single-unit) flips in the initial condition $\ket{\xi_1}$ relaxing at equilibrium with a large enough $m_1$ to guarantee recall.
Note that, since $\Phi(x)$ is an odd function, also the anti-pattern $-\ket{\xi_1}$ has the related recall state.
This symmetry imposes a theoretical upper bound of $\delta = 0.5$.
As shown in \FigRef{fig:Diagram}(f), this maximum $\delta$ is approached at $T = 0$ ($g\to\infty$), demonstrating that the newly found recall states possess exceptionally wide basins of attraction.
Even as synaptic strength $g$ decreases (equivalent to increasing temperature $T$) in the regions where recall states persist, the RAS$_1$ basins exhibit only moderate shrinkage.
Remarkably, $\delta$ shows weak dependence of $\alpha$ in the recall phase, suggesting a robust decoupling between memory load and attractor stability.

\section{Coexistence of recall and glassy states}

\begin{figure}[!htb]
   \centering
   \includegraphics[width=86mm]{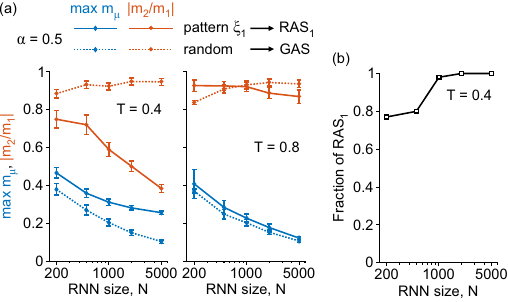}
   \caption{Coexistence of RAS$_1$ and glassy states. 
	(a) Maximum overlap $\max |m_\mu|$ (light blue) and ratio between second and first largest overlaps $|m_2/m_1|$ (brown) at the equilibrium state where the networks relax.
	Two different initial conditions are imposed: $\ket{\xi_1}$ (solid) and a random linear combination of all patterns (dashed).  
	For each network size $N$ and condition, averages are over 100 networks with randomly chosen patterns and initial conditions.
	Error bars, 3 SEM.
   Memory load $\alpha=0.5$.
	Left and right panels, have $T = 1/g = 0.4$ and $0.8$, respectively.
	(b) Fraction of equilibria found in the 100 simulations with initial condition $\ket{\xi_1}$ where the pattern is recalled (i.e., $|m_1| > 5 \sqrt{v/N}$) at $T = 0.4$ and varying $N$.}
   \label{fig:RASandGAS}
\end{figure}

So far we proved the persistence of pure recall states with a wide basin of attraction also beyond the critical $\alpha_c$ inducing in spin-glass models a catastrophic forgetting.
In this widened recall phase, RAS$_1$ displays a highly symmetric minimum (small $c_v$ of the Jacobian eigenvalues $\lambda_n$).
However, when moving downward in \FigRef{fig:Diagram}(b) from the glassy phase by increasing the synaptic strength $g$ (i.e., by reducing $T$), a discontinuity in the size $v$ of the uncondensed noise is apparent only for $\alpha \lesssim 0.2$.
This is suggestive of a qualitative difference between the classical recall states found at low-temperature for $\alpha<\alpha_c$, and the new ones characterized here for relatively large memory loads.
In search for the root of such diversity, we first inspected in the large-$\alpha$ regime the coexistence of GAS and RAS$_1$ usually expected only at relatively small $\alpha$ \cite{Gardner1986,Treves1988,Amit1989}.
In numerical simulations, when initializing the network with states whose overlaps $m_\mu$ are randomly sampled from a zero-mean Gaussian distribution of small variance, the RNN dynamics consistently converge to a GAS (if it exists).
This is clearly the case at $\alpha = 0.5$ and $T = 0.8$ ($g = 1.25$) shown in \FigRef{fig:RASandGAS}(a)-right.
Indeed, at increasing $N$ the maximum overlap found tends to zero and the first and second greatest overlaps have almost the same absolute value: no patterns can be distinguished.   
The same happens if $\ket{r(0)} = \xi_1$ (solid lines), confirming that in this phase pure recall states are absent.
However, when the coupling strength $g$ is increased (effectively reducing $T$ to $0.4$) pushing RNNs with self-couplings into the new recall phase, the same two initial conditions now determine a divergence into two distinct attractors [\FigRef{fig:RASandGAS}(a)-left].
Starting from the first stored pattern a RAS$_1$ is approached as $m_1$ displays i) a marked separation with the second largest overlap ($|m_2/m_1|\to 0$) and ii) a non-zero asymptotic value.
Furthermore, by increasing $N$ the likelihood to fall into a pure recall state approaches 1 [\FigRef{fig:RASandGAS}(b)].
Relaxation from random initial conditions is instead indistinguishable from the one observed in the glassy phase ($T = 0.8$), proving the coexistence of GAS and RAS$_1$ for sufficiently low $T$ and $\alpha > \alpha_c$.

\begin{figure}[!htb]
   \centering
   \includegraphics[width=86mm]{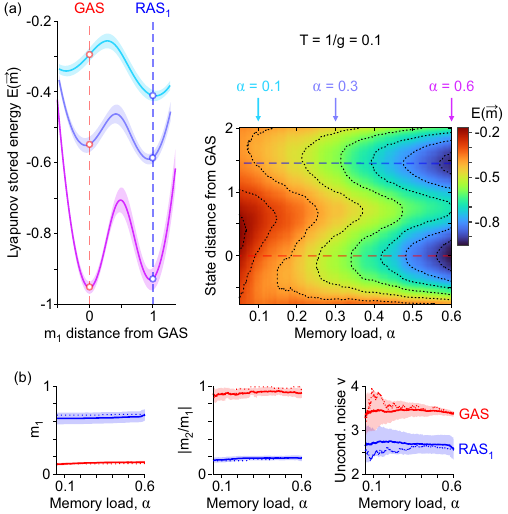}
   \caption{RAS$_1$-GAS energy landscape at varying $\alpha$. 
   (a) For each memory load, Lyapunov  functions $E(\vec{m})$ are evaluated on the linear manifold linking the found RAS$_1$ and GAS.
   Averages and standard deviations (SD, shaded strips) are computed over 20 RNNs ($N = 2000$, $g = 10$) with independently sampled random patterns.
   The position on the manifold is given by the relative overlaps $(m_1-m_1^\mathrm{GAS})/(m_1^\mathrm{RAS}-m_1^\mathrm{GAS})$ (left) and the normalized distance $||\ket{r} - \ket{r^\mathrm{GAS}}||/||\ket{r^\mathrm{RAS}} - \ket{r^\mathrm{GAS}}||$ (right) from the found GAS.
   Dashed lines: red and blue, position of the found GAS and RAS$_1$, respectively.
   In each RNN, the first (at $\alpha = 0.6$) RAS$_1$ and GAS is computed as in \FigRef{fig:RASandGAS}(a).
   Then $\alpha$ is reduced by removing one pattern per time in $\mathbf{J}$. 
   New RAS$_1$ and GAS are then found from the relaxation dynamics starting from the equilibria of the previous $\alpha$.
   (b) Left: overlaps $m_1$ with the first pattern used as initial condition at $\alpha = 0.6$ to find RAS$_1$, and never removed.
   Middle: ratio $|m_2/m_1|$ as in \FigRef{fig:RASandGAS}(a).
   Right: size $v$ of the uncondensed noise.
   Mean (solid lines) and SD (shaded strip) across the simulated RNN in (a), in the found GAS (red) and RAS$_1$ (blue).
   Dotted lines, same observables for an example RNN among the 20 simulated.}
   \label{fig:RASGASlandscape}
\end{figure}

This marks another deep difference with what observed in spin-glass models of associative memories. 
Self-couplings widen the pure recall phase without shrinking the glassy phase. 
Consequently, when glassy states emerge as metastable states at relatively low temperatures, increasing the number $P$ of stored patterns should preserve the coarse-grained structure of the energy landscape.
In \FigRef{fig:RASGASlandscape}(a), we show this is exactly the case for sufficiently large coupling strengths $g$ ($T = 0.1$). 
Specifically, by tracking the position of the pure recall and glassy equilibria found at $\alpha = 0.5$ as stored patterns are progressively removed (i.e., as $\alpha$ decreases), the sections of the energy landscape consistently exhibit bistability.
Note that GAS and RAS$_1$ found not always are located at the bottom of the corresponding valleys [\FigRef{fig:RASGASlandscape}(a)-left].
This is the evidence that they are local minima constellating the two macroscopic basins of attractions associated to a less energetic GAS and RAS$_1$, respectively.
These local minima does not change their characterizing properties when the stored patterns are removed: GASs display always large uncondensed noise and small overlaps, while in RAS$_1$s the overlap $m_1$ is consistently separable from all the $m_{\mu>1}$ [\FigRef{fig:RASGASlandscape}(b)].   
The roughness of the energy landscape \cite{Amit1989} is then a preserved feature also when self-couplings are taken into account.
However, a notable distinction arises: in the recall phase of spin-glass models, RAS$_1$s exhibit higher energy than GASs when memory load $\alpha \lesssim \alpha_c$.
In our RNNs with self-couplings the opposite is true [see the case $\alpha = 0.1$ in \FigRef{fig:RASGASlandscape}(a)].
Glassy states are then less-favorable network states at large $g$, although their energy $E(\vec{m})$ lowers with $\alpha$ reaching the same level of RAS$_1$ at $\alpha = 0.5$.

\section{Mixture states beyond $\alpha_c$}

\begin{figure}[!htb]
   \centering 
   \includegraphics[width=86mm]{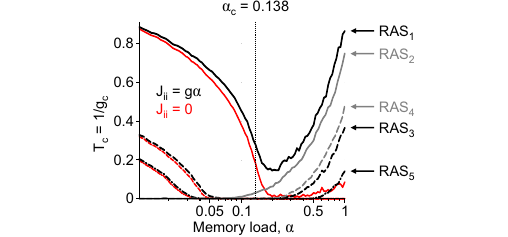}
   \caption{Boundary $T_c = 1/g_c$ of recall pure (RAS$_1$) and mixture (RAS$_{n>1}$) states. 
	Recall attractor states RAS$_n$ are the memories associated with mixtures $\sgn{\left(\sum_{\nu =1}^{n} \ket{\xi_\mu}\right)}$. 
	In the numerical integration, the equilibrium where the RNN relaxes starting from this initial condition, is a RAS$_n$ if the $|m_\mu| > 5 \sqrt{v/N}$ for all $\mu<n$.
	Boundaries are averages $g_c$ over $100$ random networks with $N=500$. 
	For a given number $P$ of stored patterns, $g_c$ is computed in each network by decreasing $g$ from 1000 (i.e., $T=10^-3$) until the related RAS is no more recalled.
	Black and gray lines are the boundaries for the intact $P$-rank $\mathbf{J}$ (i.e., $J_{ii}=g \alpha$), while the red ones are obtained for the same RNNs by forcing $J_{ii} = 0$ (i.e., spin-glass models). 
	Dotted line, limit capacity $\alpha_c = 0.138$ for spin-glass models \cite{Amit1985}.}
    \label{fig:Mixtures}
\end{figure}

The observed coexistence of glassy and pure recall states at memory loads exceeding $\alpha_c$ demonstrates the stabilizing role of self-couplings in enabling associative memory implementations for stored patterns.
In analogy with standard spin-glass models, which feature both GAS and RAS$_1$, there exists a multitude of local minima often referred to as `mixtures' \cite{Amit1989}. This raises a natural question: do these mixtures also become further stabilized when $J_{ii} = g \alpha$?
In networks of spin-like neurons, any state given by a symmetric linear combination of an odd number of patterns, such as $\ket{r}=\sgn(\sum_{\mu=1}^{2*k+1} \ket{\xi_{\mu}})$, is stable in the limit $N\to\infty$ as long as $\alpha \to 0$ \cite{Amit1989}. 
In \FigRef{fig:LyapunovFunc}(c) we have shown for small $\alpha$ that these odd mixtures are stable states also in our case.

Here we further investigate the stability of these mixtures in simulations by applying the RAS$_1$ recall criterion (i.e., $|m_\nu|>5\sqrt{v/N}$ for all pattern $\mu$s composing the mixture) and computing the critical lines $T_c(\alpha) = 1/g_c(\alpha)$ above which they lose stability (\FigRef{fig:Mixtures}, black lines). 
As expected, these critical lines consistently occur at lower $T_c$ values for the same RNNs but without self-couplings (red lines), further highlighting the stabilizing effect of autapses.
We found that, unlike the recall states, there exists an interval of $\alpha$ values for which self-couplings fail to stabilize these states.
In line with standard spin-glass models, odd mixtures lose stability at progressively smaller critical memory loads compared to $\alpha_c$ as the number of composing patterns increases. 
However, as $\alpha$ increases beyond approximately $0.3$, RAS$_3$ can once again be recalled, and higher memory loads progressively enable the retrieval of mixtures with larger numbers of patterns.

Notably, this behavior is absent in standard spin-glass models without self-couplings.
A striking difference emerges in RNNs with autapses: even mixtures can be faithfully recalled at sufficiently large $\alpha$ (\FigRef{fig:Mixtures}, gray lines). 
This reveals a cornucopia of novel attractor states that were previously unattainable in the absence of self-couplings.

\section{Discussion}

Here we studied the associative memory properties of deterministic RNNs with an Amari-Hopfield synaptic matrix \cite{Amari1972,Hopfield1982}. 
Unlike the approach often used in spin-glass models, we did not set the diagonal elements of the matrix to zero. 
Rather we strictly adopted the Hebbian learning rule, which is local and naturally includes self-couplings ($J_{ii} > 0$).
The main result we report is the capability of such RNNs to encode associative memories of $P$ stored patterns which is proportional to the number $N$ of units composing the networks (i.e., constant $\alpha = P/N$ also in the limit $N\to\infty$).
This well beyond the critical memory load $\alpha_c \simeq 0.138$ found in spin glass models \cite{Amit1985,Kuhn1991}, that lowers as temperature is increased ($T>0$).
Self-couplings have a stabilizing role as they introduce an additional state-dependent field (i.e., a synaptic current) further pushing the network in the direction of the approached equilibria.
As a result, not only pure recall states regain stability at high memory load, but also a multitude of other attractor states coexist in the same region of the phase diagram.
This introduces a certain roughness to the basins of attraction of RAS$_1$, which, from a computational perspective, presents a valuable opportunity. 
Indeed, beyond its capacity to retain an associative memory when the system settles into the related basin of attraction, it can continue transitioning between local and correlated minima --  effectively functioning as algorithmic steps in an additional, context-dependent computation.

The presence of self-couplings in RNNs with spin-like discrete and stochastic neurons presents a significant challenge. 
This is particularly true in equilibrium statistical mechanics, where the lack of a conventional free energy function complicates ensuring the system's stability \cite{Kanter1987,Amit1989}.
This absence means that the network dynamics cannot be easily understood or predicted using traditional tools from statistical physics complicating the analysis and requiring the use of other theoretical approaches.
In response to the challenges posed by self-couplings, the theoretical framework applied in \cite{Singh2001,Coolen1993,Coolen1994} provides a robust method for analyzing the stochastic dynamics of RNNs with self-interactions. 
This approach has proven effective in providing insights into the behavior of spin-glass models of neural networks, offering a new perspective on how these systems can be stabilized and controlled despite the absence of a conventional free energy description.

The impact of self-couplings in RNNs has been a subject of extensive debate and research. 
Early studies suggested that they might destabilize the recall states or have negligible effects \cite{Kuhn1991,Fontanari1988}. 
In particular, in \cite{Kanter1987} the authors argued that despite the fact that autapses make recall stable even beyond $\alpha_c$ their basin attraction shrinks to 0 for large $\alpha$. 
Other works \cite{Krauth1988,Singh2001} provided observations that seem going in the opposite direction. 
With our findings in RNNs, we further shed some light on this open issue and sharpen the gap between model and biology.
Indeed, in this framework single-units aim at modeling the collective firing rate of a local population of spiking neurons \cite{Abbott1993,Treves1993,Brunel1999,Knight2000,Mattia2002}.
Keeping closer to biology allowed us to find that self-couplings widen the basins of attraction of the recall states. 
This is only one of the advantages in incorporating autapses in associative network models, in fact we found the whole energy landscape (i.e., the Lyapunov function) is radically different from the free energy of spin-glass models. 
For example, even symmetric mixtures of patterns are stabilized for large enough $\alpha$.

Over the years, various adjustments to the synaptic matrix have been proposed to address the issue of catastrophic forgetting in spin-glass models. 
One of the first successful approaches introduced a nonlocal method for synaptic learning, resulting in an optimal projection matrix that ensures stored patterns remain attractor states, provided they are independent \cite{Personnaz1985,Kanter1987}. 
More or less at the same time, a `dreaming' procedure interpreted as the `night' phase where spurious memories are erased, has been introduced \cite{Hopfield1983}.
It is based on the evidence that spurious states are exponentially more abundant than the number of pure recall states, in particular when the memory load is close to the critical value $\alpha > 0.14$ (they focused on the case $T=0$).
Although this approach relies solely on local information to modify the elements of the synaptic matrix, it does come with certain limitations. 
One such drawback concerns the extent of unlearning performed. 
In particular, excessive unlearning can erode learned memory (coming back to catastrophic forgetting), as these states increasingly become the more stable equilibria.
A promising improvement has been proposed to mitigate this issue \cite{Fachechi2019}, alongside more recent advancements \cite{Agliari2024,Serricchio2025}. 
In this refined approach, unlearning is complemented by a consolidation process, allowing the system to asymptotically converge toward the optimal projection matrix described in \cite{Personnaz1985,Kanter1987}.
While effective, all these strategies require a costly reorganization of the synaptic matrix, which in this work we demonstrate it is unnecessary.

Finally, it is important to remark that our theoretical framework remains exact even for a finite network size $N$. 
Notably, in the thermodynamic limit ($N \to \infty$), the dynamics \eqref{eq:MDyn} of the overlaps $\vec{m}$ align with that observed for discrete neurons undergoing stochastic sequential dynamics \cite{Coolen2001b}. 
This suggests that our conclusions may have a rather universal character.

\begin{acknowledgments}
We thank G. Mongillo, A. Treves, L. Mazzucato, S. Suweis and A. Maritan for stimulating discussions.
Work partially funded by the Italian National Recovery and Resilience Plan (PNRR), M4C2, funded by the European Union - NextGenerationEU (Project IR0000011, CUP B51E22000150006, ‘EBRAINS-Italy’) and (Project PE00000006, CUP I83C24000990005 ‘MNESYS’) to M.M.
\end{acknowledgments}

\appendix

\section{Derivation of Lyapunov function $E(\vec{m})$}
\label{app:LyapFunc}

By definition the Lyapunov function in pattern subspace has 
\begin{displaymath}
	\frac{dE}{dt} = \sum_{\mu=1}^P \partial_{m_\mu}E \, \dot{m}_\mu \leq 0 \, .
\end{displaymath}
To satisfy this inequality, it is sufficient to request that
\begin{displaymath}
	\partial_{m_\mu}E = - \tau \dot{m}_\mu \, ,
\end{displaymath}
as it leads in the r.h.s. of the above inequality to have a sum of non-positive terms $-\tau \dot{m}_\mu^2$.
Integrating this ordinary differential equation (ODE) by taking into account \EqRef{eq:SingleMDyn}, we eventually obtain:
\begin{displaymath}
	E(\vec{m}) = \frac{1}{2} m_\mu^2 
              - \frac{1}{g N} \sum_{i = 1}^N F\left(g \sum_{\nu = 1}^P\xi_{\nu i} m_\nu \right) 
              + G_\mu(\{m_\nu\}_{\nu \neq \mu})\, ,
\end{displaymath}
where $F(h) = \int \Phi(h) dh$.
$G_\mu$ is a function not-depending on $m_\mu$ to be determined by deriving the Lyapunov function with respect to $m_\sigma$ ($\sigma \neq \mu$):
\begin{displaymath}
\begin{split}
	\partial_{m_\sigma} E =& - \frac{1}{N} \sum_{i = 1}^N \xi_{\sigma i} \Phi\left(g \sum_{\nu = 1}^P\xi_{\nu i} m_\nu \right) + \partial_{m_\sigma} G_\mu \\
   =& -\braket{\xi_\sigma|\Phi(\vec{m})} + \partial_{m_\sigma} G_\mu
\end{split} \, .
\end{displaymath}
Considering that the value of the inner product is given by \EqRef{eq:SingleMDyn}, and that for the $\sigma$ pattern $\partial_{m_\sigma}E = - \tau \dot{m}_\sigma$, 
\begin{displaymath}
	\partial_{m_\sigma} G_\mu = m_\sigma \, .
\end{displaymath}
Solving this ODE for any $\sigma \neq \mu$ leads to $G_\mu = \frac{1}{2} \sum_{\sigma \neq \mu} m_\sigma^2$, and eventually to \EqRef{eq:LyapunovFunc}.

\section{Theoretical predictions for the stationary overlaps}
\label{app:LowAlphaSCE}

We are searching the solution with the dominant pattern being the first one for convenience.
We start by separating the uncondensed part of the current from the rest:
\begin{displaymath}
	\ket{h} = g \left( m_1 \ket{\xi_1} + \sum_{\mu=2}^P m_\mu \ket{\xi_\mu} \right)
\end{displaymath}
with $P=\alpha N$. 
Following \EqRef{eq:FixedPoints} for arbitrary $m_1$ we use as ansatz:
\begin{displaymath}
   m_{\mu}=m_1 \left\{\mathbf{\Xi}^+ \ket{\xi_1}\right\}_\mu
\end{displaymath}
which allows us the rewrite the current as:
\begin{displaymath}
   h_{i} = g m_1 \xi_{1 i} \left[1 + \frac{P-1}{N} + \xi_{1 i} h^{(\mathrm{uc})}_i  \right] \, ,
\end{displaymath}
where the uncondensed part of the current is 
\begin{displaymath}
	h^{(\mathrm{uc})}_i = \frac{1}{N}\sum_{\mu>1}^{P} \sum_{j\neq i}^{N} \xi_{\mu i}\xi_{\mu j} \xi_{1 j} \, .
\end{displaymath}
In the large $N$ limit this part of the current is equivalent to a Gaussian process with zero mean and variance $\mathbb{V} \left[ h^{(\mathrm{uc})}\right ]=v \alpha $ where $v$ is the order parameter in \EqRef{eq:UncondNoiseVar}. 
The overlap $m_1$ can thus be rewritten as:
\begin{displaymath}
	m_1= \frac{1}{N}\sum_{i=1}^{N} \xi_{1 i} \Phi(g m_1 \xi_{1 i} (1 +\alpha \sqrt{v \alpha} z_i) \, ,
\end{displaymath}
where the $z_i$ are independent random variables sampled from a Gaussian distribution with 0 mean and unitary variance. 
From here the self-consistency equation is solved resorting to the mean-field theory for spin-glass models \cite{Amit1989}.
As $\Phi(x) = \tanh(x)$ from \EqRef{eq:Tanh} is an odd function, in the thermodynamic limit the sum can be replaced by the following integral equation:
\begin{displaymath}
	m_1 =\int_{-\infty}^{\infty} \tanh(g m_1 z) d\mu_{\alpha,v}(z) \, ,
\end{displaymath} 
where $d\mu_{\alpha,v}=e^{\frac{(z-\alpha -1)^2}{2 v \alpha}}dz/\sqrt{2 \pi v \alpha}$. 
An analogous equation for the order parameter $q =\langle \langle \sum_{i=1}^{N} r_i^2 /N \rangle \rangle$ is obtained in the same way. 
Here the double angular brackets indicates the average over the quenched variables \cite{Amit1989}. 
If we take the limit $g \to \infty$ (i.e., $T \to 0$) and use either $v=1$ or the $v$ from the replica ansatz, the two expressions for $m_1$ in the main text are recovered.

\bibliography{RefLibrary}

\providecommand{\noopsort}[1]{}\providecommand{\singleletter}[1]{#1}%
\begin{thebibliography}{47}%
\makeatletter
\providecommand \@ifxundefined [1]{%
 \@ifx{#1\undefined}
}%
\providecommand \@ifnum [1]{%
 \ifnum #1\expandafter \@firstoftwo
 \else \expandafter \@secondoftwo
 \fi
}%
\providecommand \@ifx [1]{%
 \ifx #1\expandafter \@firstoftwo
 \else \expandafter \@secondoftwo
 \fi
}%
\providecommand \natexlab [1]{#1}%
\providecommand \enquote  [1]{``#1''}%
\providecommand \bibnamefont  [1]{#1}%
\providecommand \bibfnamefont [1]{#1}%
\providecommand \citenamefont [1]{#1}%
\providecommand \href@noop [0]{\@secondoftwo}%
\providecommand \href [0]{\begingroup \@sanitize@url \@href}%
\providecommand \@href[1]{\@@startlink{#1}\@@href}%
\providecommand \@@href[1]{\endgroup#1\@@endlink}%
\providecommand \@sanitize@url [0]{\catcode `\\12\catcode `\$12\catcode
  `\&12\catcode `\#12\catcode `\^12\catcode `\_12\catcode `\%12\relax}%
\providecommand \@@startlink[1]{}%
\providecommand \@@endlink[0]{}%
\providecommand \url  [0]{\begingroup\@sanitize@url \@url }%
\providecommand \@url [1]{\endgroup\@href {#1}{\urlprefix }}%
\providecommand \urlprefix  [0]{URL }%
\providecommand \Eprint [0]{\href }%
\providecommand \doibase [0]{https://doi.org/}%
\providecommand \selectlanguage [0]{\@gobble}%
\providecommand \bibinfo  [0]{\@secondoftwo}%
\providecommand \bibfield  [0]{\@secondoftwo}%
\providecommand \translation [1]{[#1]}%
\providecommand \BibitemOpen [0]{}%
\providecommand \bibitemStop [0]{}%
\providecommand \bibitemNoStop [0]{.\EOS\space}%
\providecommand \EOS [0]{\spacefactor3000\relax}%
\providecommand \BibitemShut  [1]{\csname bibitem#1\endcsname}%
\let\auto@bib@innerbib\@empty
\bibitem [{\citenamefont {Hebb}(1949)}]{Hebb1949}%
  \BibitemOpen
  \bibfield  {author} {\bibinfo {author} {\bibfnamefont {D.~O.}\ \bibnamefont
  {Hebb}},\ }\href@noop {} {\emph {\bibinfo {title} {{The Organization of
  Behavior: A Neuropsychological Theory}}}}\ (\bibinfo  {publisher} {John Wiley
  \& Sons, Inc.},\ \bibinfo {address} {New York},\ \bibinfo {year} {1949})\ p.\
  \bibinfo {pages} {335}\BibitemShut {NoStop}%
\bibitem [{\citenamefont {Magee}\ and\ \citenamefont
  {Grienberger}(2020)}]{Magee2020}%
  \BibitemOpen
  \bibfield  {author} {\bibinfo {author} {\bibfnamefont {J.~C.}\ \bibnamefont
  {Magee}}\ and\ \bibinfo {author} {\bibfnamefont {C.}~\bibnamefont
  {Grienberger}},\ }\href {https://doi.org/10.1146/annurev-neuro-090919-022842}
  {\bibfield  {journal} {\bibinfo  {journal} {Annu.\ Rev.\ Neurosci.}\ }\textbf
  {\bibinfo {volume} {43}},\ \bibinfo {pages} {95} (\bibinfo {year}
  {2020})}\BibitemShut {NoStop}%
\bibitem [{\citenamefont {Carrillo-Reid}\ \emph {et~al.}(2016)\citenamefont
  {Carrillo-Reid}, \citenamefont {Yang}, \citenamefont {Bando}, \citenamefont
  {Peterka},\ and\ \citenamefont {Yuste}}]{CarrilloReid2016}%
  \BibitemOpen
  \bibfield  {author} {\bibinfo {author} {\bibfnamefont {L.}~\bibnamefont
  {Carrillo-Reid}}, \bibinfo {author} {\bibfnamefont {W.}~\bibnamefont {Yang}},
  \bibinfo {author} {\bibfnamefont {Y.}~\bibnamefont {Bando}}, \bibinfo
  {author} {\bibfnamefont {D.~S.}\ \bibnamefont {Peterka}},\ and\ \bibinfo
  {author} {\bibfnamefont {R.}~\bibnamefont {Yuste}},\ }\href
  {https://doi.org/10.1126/science.aaf7560} {\bibfield  {journal} {\bibinfo
  {journal} {Science}\ }\textbf {\bibinfo {volume} {353}},\ \bibinfo {pages}
  {691} (\bibinfo {year} {2016})}\BibitemShut {NoStop}%
\bibitem [{\citenamefont {Amit}(1989)}]{Amit1989}%
  \BibitemOpen
  \bibfield  {author} {\bibinfo {author} {\bibfnamefont {D.~J.}\ \bibnamefont
  {Amit}},\ }\href {https://doi.org/10.1017/CBO9780511623257} {\emph {\bibinfo
  {title} {{Modeling brain function}}}}\ (\bibinfo  {publisher} {Cambridge
  University Press},\ \bibinfo {year} {1989})\ p.\ \bibinfo {pages}
  {504}\BibitemShut {NoStop}%
\bibitem [{\citenamefont {Amari}(1972)}]{Amari1972}%
  \BibitemOpen
  \bibfield  {author} {\bibinfo {author} {\bibfnamefont {S.-i.}\ \bibnamefont
  {Amari}},\ }\href {https://doi.org/10.1109/T-C.1972.223477} {\bibfield
  {journal} {\bibinfo  {journal} {IEEE Trans.\ Comput.}\ }\textbf {\bibinfo
  {volume} {C-21}},\ \bibinfo {pages} {1197} (\bibinfo {year}
  {1972})}\BibitemShut {NoStop}%
\bibitem [{\citenamefont {Hopfield}(1982)}]{Hopfield1982}%
  \BibitemOpen
  \bibfield  {author} {\bibinfo {author} {\bibfnamefont {J.~J.}\ \bibnamefont
  {Hopfield}},\ }\href {https://doi.org/10.1073/pnas.79.8.2554} {\bibfield
  {journal} {\bibinfo  {journal} {Proc.\ Natl.\ Acad.\ Sci.\ USA}\ }\textbf
  {\bibinfo {volume} {79}},\ \bibinfo {pages} {2554} (\bibinfo {year}
  {1982})}\BibitemShut {NoStop}%
\bibitem [{\citenamefont {Miyashita}(1988)}]{Miyashita1988}%
  \BibitemOpen
  \bibfield  {author} {\bibinfo {author} {\bibfnamefont {Y.}~\bibnamefont
  {Miyashita}},\ }\href {https://doi.org/10.1038/335817a0} {\bibfield
  {journal} {\bibinfo  {journal} {Nature}\ }\textbf {\bibinfo {volume} {335}},\
  \bibinfo {pages} {817} (\bibinfo {year} {1988})}\BibitemShut {NoStop}%
\bibitem [{\citenamefont {Amit}(1995)}]{Amit1995}%
  \BibitemOpen
  \bibfield  {author} {\bibinfo {author} {\bibfnamefont {D.~J.}\ \bibnamefont
  {Amit}},\ }\href@noop {} {\bibfield  {journal} {\bibinfo  {journal} {Behav.\
  Brain Sci.}\ }\textbf {\bibinfo {volume} {18}},\ \bibinfo {pages} {617}
  (\bibinfo {year} {1995})}\BibitemShut {NoStop}%
\bibitem [{\citenamefont {Hertz}\ \emph {et~al.}(1991)\citenamefont {Hertz},
  \citenamefont {Krogh},\ and\ \citenamefont {Palmer}}]{Hertz1991}%
  \BibitemOpen
  \bibfield  {author} {\bibinfo {author} {\bibfnamefont {J.~A.}\ \bibnamefont
  {Hertz}}, \bibinfo {author} {\bibfnamefont {A.}~\bibnamefont {Krogh}},\ and\
  \bibinfo {author} {\bibfnamefont {R.~G.}\ \bibnamefont {Palmer}},\
  }\href@noop {} {\emph {\bibinfo {title} {{Introduction to the theory of
  neural computation}}}}\ (\bibinfo  {publisher} {West view Press},\ \bibinfo
  {address} {Boca Raton},\ \bibinfo {year} {1991})\ p.\ \bibinfo {pages}
  {327}\BibitemShut {NoStop}%
\bibitem [{\citenamefont {Song}\ \emph {et~al.}(2005)\citenamefont {Song},
  \citenamefont {Sj{\"{o}}str{\"{o}}m}, \citenamefont {Reigl}, \citenamefont
  {Nelson},\ and\ \citenamefont {Chklovskii}}]{Song2005}%
  \BibitemOpen
  \bibfield  {author} {\bibinfo {author} {\bibfnamefont {S.}~\bibnamefont
  {Song}}, \bibinfo {author} {\bibfnamefont {P.~J.}\ \bibnamefont
  {Sj{\"{o}}str{\"{o}}m}}, \bibinfo {author} {\bibfnamefont {M.}~\bibnamefont
  {Reigl}}, \bibinfo {author} {\bibfnamefont {S.~B.}\ \bibnamefont {Nelson}},\
  and\ \bibinfo {author} {\bibfnamefont {D.~B.}\ \bibnamefont {Chklovskii}},\
  }\href {https://doi.org/10.1371/journal.pbio.0030068} {\bibfield  {journal}
  {\bibinfo  {journal} {PLoS Biol.}\ }\textbf {\bibinfo {volume} {3}},\
  \bibinfo {pages} {e68} (\bibinfo {year} {2005})}\BibitemShut {NoStop}%
\bibitem [{\citenamefont {Lefort}\ \emph {et~al.}(2009)\citenamefont {Lefort},
  \citenamefont {Tomm}, \citenamefont {{Floyd Sarria}},\ and\ \citenamefont
  {Petersen}}]{Lefort2009}%
  \BibitemOpen
  \bibfield  {author} {\bibinfo {author} {\bibfnamefont {S.}~\bibnamefont
  {Lefort}}, \bibinfo {author} {\bibfnamefont {C.}~\bibnamefont {Tomm}},
  \bibinfo {author} {\bibfnamefont {J.-C.}\ \bibnamefont {{Floyd Sarria}}},\
  and\ \bibinfo {author} {\bibfnamefont {C.~C.~H.}\ \bibnamefont {Petersen}},\
  }\href {https://doi.org/10.1016/j.neuron.2008.12.020} {\bibfield  {journal}
  {\bibinfo  {journal} {Neuron}\ }\textbf {\bibinfo {volume} {61}},\ \bibinfo
  {pages} {301} (\bibinfo {year} {2009})}\BibitemShut {NoStop}%
\bibitem [{\citenamefont {Coolen}\ \emph {et~al.}(2005)\citenamefont {Coolen},
  \citenamefont {K{\"{u}}hn},\ and\ \citenamefont {Sollich}}]{Coolen2005}%
  \BibitemOpen
  \bibfield  {author} {\bibinfo {author} {\bibfnamefont {A.~C.~C.}\
  \bibnamefont {Coolen}}, \bibinfo {author} {\bibfnamefont {R.}~\bibnamefont
  {K{\"{u}}hn}},\ and\ \bibinfo {author} {\bibfnamefont {P.}~\bibnamefont
  {Sollich}},\ }\href@noop {} {\emph {\bibinfo {title} {{Theory of Neural
  Information Processing Systems}}}}\ (\bibinfo  {publisher} {Oxford University
  Press},\ \bibinfo {address} {New York},\ \bibinfo {year} {2005})\ p.\
  \bibinfo {pages} {569}\BibitemShut {NoStop}%
\bibitem [{\citenamefont {Kanter}\ and\ \citenamefont
  {Sompolinsky}(1987)}]{Kanter1987}%
  \BibitemOpen
  \bibfield  {author} {\bibinfo {author} {\bibfnamefont {I.}~\bibnamefont
  {Kanter}}\ and\ \bibinfo {author} {\bibfnamefont {H.}~\bibnamefont
  {Sompolinsky}},\ }\href {https://doi.org/10.1103/PhysRevA.35.380} {\bibfield
  {journal} {\bibinfo  {journal} {Phys.\ Rev.\ A}\ }\textbf {\bibinfo {volume}
  {35}},\ \bibinfo {pages} {380} (\bibinfo {year} {1987})}\BibitemShut
  {NoStop}%
\bibitem [{\citenamefont {Clark}\ and\ \citenamefont
  {Abbott}(2024)}]{Clark2024}%
  \BibitemOpen
  \bibfield  {author} {\bibinfo {author} {\bibfnamefont {D.~G.}\ \bibnamefont
  {Clark}}\ and\ \bibinfo {author} {\bibfnamefont {L.~F.}\ \bibnamefont
  {Abbott}},\ }\href {https://doi.org/10.1103/PhysRevX.14.021001} {\bibfield
  {journal} {\bibinfo  {journal} {Phys.\ Rev.\ X}\ }\textbf {\bibinfo {volume}
  {14}},\ \bibinfo {pages} {021001} (\bibinfo {year} {2024})}\BibitemShut
  {NoStop}%
\bibitem [{\citenamefont {Stern}\ \emph {et~al.}(2014)\citenamefont {Stern},
  \citenamefont {Sompolinsky},\ and\ \citenamefont {Abbott}}]{Stern2014}%
  \BibitemOpen
  \bibfield  {author} {\bibinfo {author} {\bibfnamefont {M.}~\bibnamefont
  {Stern}}, \bibinfo {author} {\bibfnamefont {H.}~\bibnamefont {Sompolinsky}},\
  and\ \bibinfo {author} {\bibfnamefont {L.~F.}\ \bibnamefont {Abbott}},\
  }\href {https://doi.org/10.1103/PhysRevE.90.062710} {\bibfield  {journal}
  {\bibinfo  {journal} {Phys.\ Rev.\ E}\ }\textbf {\bibinfo {volume} {90}},\
  \bibinfo {pages} {062710} (\bibinfo {year} {2014})}\BibitemShut {NoStop}%
\bibitem [{\citenamefont {Stern}\ \emph {et~al.}(2023)\citenamefont {Stern},
  \citenamefont {Istrate},\ and\ \citenamefont {Mazzucato}}]{Stern2023}%
  \BibitemOpen
  \bibfield  {author} {\bibinfo {author} {\bibfnamefont {M.}~\bibnamefont
  {Stern}}, \bibinfo {author} {\bibfnamefont {N.}~\bibnamefont {Istrate}},\
  and\ \bibinfo {author} {\bibfnamefont {L.}~\bibnamefont {Mazzucato}},\
  }\bibfield  {journal} {\bibinfo  {journal} {eLife}\ }\textbf {\bibinfo
  {volume} {12}},\ \href {https://doi.org/10.7554/eLife.86552}
  {10.7554/eLife.86552} (\bibinfo {year} {2023})\BibitemShut {NoStop}%
\bibitem [{\citenamefont {Amit}\ \emph
  {et~al.}(1985{\natexlab{a}})\citenamefont {Amit}, \citenamefont {Gutfreund},\
  and\ \citenamefont {Sompolinsky}}]{Amit1985}%
  \BibitemOpen
  \bibfield  {author} {\bibinfo {author} {\bibfnamefont {D.~J.}\ \bibnamefont
  {Amit}}, \bibinfo {author} {\bibfnamefont {H.}~\bibnamefont {Gutfreund}},\
  and\ \bibinfo {author} {\bibfnamefont {H.}~\bibnamefont {Sompolinsky}},\
  }\href {https://doi.org/10.1103/PhysRevLett.55.1530} {\bibfield  {journal}
  {\bibinfo  {journal} {Phys.\ Rev.\ Lett.}\ }\textbf {\bibinfo {volume}
  {55}},\ \bibinfo {pages} {1530} (\bibinfo {year}
  {1985}{\natexlab{a}})}\BibitemShut {NoStop}%
\bibitem [{\citenamefont {Crisanti}\ \emph {et~al.}(1986)\citenamefont
  {Crisanti}, \citenamefont {Amit},\ and\ \citenamefont
  {Gutfreund}}]{Crisanti1986}%
  \BibitemOpen
  \bibfield  {author} {\bibinfo {author} {\bibfnamefont {A.}~\bibnamefont
  {Crisanti}}, \bibinfo {author} {\bibfnamefont {D.~J.}\ \bibnamefont {Amit}},\
  and\ \bibinfo {author} {\bibfnamefont {H.}~\bibnamefont {Gutfreund}},\ }\href
  {https://doi.org/10.1209/0295-5075/2/4/012} {\bibfield  {journal} {\bibinfo
  {journal} {Europhys.\ Lett.}\ }\textbf {\bibinfo {volume} {2}},\ \bibinfo
  {pages} {337} (\bibinfo {year} {1986})}\BibitemShut {NoStop}%
\bibitem [{\citenamefont {Personnaz}\ \emph {et~al.}(1985)\citenamefont
  {Personnaz}, \citenamefont {Guyon},\ and\ \citenamefont
  {Dreyfus}}]{Personnaz1985}%
  \BibitemOpen
  \bibfield  {author} {\bibinfo {author} {\bibfnamefont {L.}~\bibnamefont
  {Personnaz}}, \bibinfo {author} {\bibfnamefont {I.}~\bibnamefont {Guyon}},\
  and\ \bibinfo {author} {\bibfnamefont {G.}~\bibnamefont {Dreyfus}},\ }\href
  {https://doi.org/10.1051/jphyslet:01985004608035900} {\bibfield  {journal}
  {\bibinfo  {journal} {J.\ Phys.\ Lett.}\ }\textbf {\bibinfo {volume} {46}},\
  \bibinfo {pages} {359} (\bibinfo {year} {1985})}\BibitemShut {NoStop}%
\bibitem [{\citenamefont {Hopfield}\ \emph {et~al.}(1983)\citenamefont
  {Hopfield}, \citenamefont {Feinstein},\ and\ \citenamefont
  {Palmer}}]{Hopfield1983}%
  \BibitemOpen
  \bibfield  {author} {\bibinfo {author} {\bibfnamefont {J.~J.}\ \bibnamefont
  {Hopfield}}, \bibinfo {author} {\bibfnamefont {D.~I.}\ \bibnamefont
  {Feinstein}},\ and\ \bibinfo {author} {\bibfnamefont {R.~G.}\ \bibnamefont
  {Palmer}},\ }\href {https://doi.org/10.1038/304158a0} {\bibfield  {journal}
  {\bibinfo  {journal} {Nature}\ }\textbf {\bibinfo {volume} {304}},\ \bibinfo
  {pages} {158} (\bibinfo {year} {1983})}\BibitemShut {NoStop}%
\bibitem [{\citenamefont {Diederich}\ and\ \citenamefont
  {Opper}(1987)}]{Diederich1987}%
  \BibitemOpen
  \bibfield  {author} {\bibinfo {author} {\bibfnamefont {S.}~\bibnamefont
  {Diederich}}\ and\ \bibinfo {author} {\bibfnamefont {M.}~\bibnamefont
  {Opper}},\ }\href {https://doi.org/10.1103/PhysRevLett.58.949} {\bibfield
  {journal} {\bibinfo  {journal} {Phys.\ Rev.\ Lett.}\ }\textbf {\bibinfo
  {volume} {58}},\ \bibinfo {pages} {949} (\bibinfo {year} {1987})}\BibitemShut
  {NoStop}%
\bibitem [{\citenamefont {Mattia}\ and\ \citenamefont {{Del
  Giudice}}(2002)}]{Mattia2002}%
  \BibitemOpen
  \bibfield  {author} {\bibinfo {author} {\bibfnamefont {M.}~\bibnamefont
  {Mattia}}\ and\ \bibinfo {author} {\bibfnamefont {P.}~\bibnamefont {{Del
  Giudice}}},\ }\href {https://doi.org/10.1103/PhysRevE.66.051917} {\bibfield
  {journal} {\bibinfo  {journal} {Phys.\ Rev.\ E}\ }\textbf {\bibinfo {volume}
  {66}},\ \bibinfo {pages} {051917} (\bibinfo {year} {2002})}\BibitemShut
  {NoStop}%
\bibitem [{\citenamefont {Abbott}\ and\ \citenamefont {van
  Vreeswijk}(1993)}]{Abbott1993}%
  \BibitemOpen
  \bibfield  {author} {\bibinfo {author} {\bibfnamefont {L.~F.}\ \bibnamefont
  {Abbott}}\ and\ \bibinfo {author} {\bibfnamefont {C.}~\bibnamefont {van
  Vreeswijk}},\ }\href@noop {} {\bibfield  {journal} {\bibinfo  {journal}
  {Phys.\ Rev.\ E}\ }\textbf {\bibinfo {volume} {48}},\ \bibinfo {pages} {1483}
  (\bibinfo {year} {1993})}\BibitemShut {NoStop}%
\bibitem [{\citenamefont {Treves}(1993)}]{Treves1993}%
  \BibitemOpen
  \bibfield  {author} {\bibinfo {author} {\bibfnamefont {A.}~\bibnamefont
  {Treves}},\ }\href@noop {} {\bibfield  {journal} {\bibinfo  {journal}
  {Network}\ }\textbf {\bibinfo {volume} {4}},\ \bibinfo {pages} {259}
  (\bibinfo {year} {1993})}\BibitemShut {NoStop}%
\bibitem [{\citenamefont {Brunel}\ and\ \citenamefont
  {Hakim}(1999)}]{Brunel1999}%
  \BibitemOpen
  \bibfield  {author} {\bibinfo {author} {\bibfnamefont {N.}~\bibnamefont
  {Brunel}}\ and\ \bibinfo {author} {\bibfnamefont {V.}~\bibnamefont {Hakim}},\
  }\href {https://doi.org/10.1162/089976699300016179} {\bibfield  {journal}
  {\bibinfo  {journal} {Neural Comput.}\ }\textbf {\bibinfo {volume} {11}},\
  \bibinfo {pages} {1621} (\bibinfo {year} {1999})}\BibitemShut {NoStop}%
\bibitem [{\citenamefont {Knight}(2000)}]{Knight2000}%
  \BibitemOpen
  \bibfield  {author} {\bibinfo {author} {\bibfnamefont {B.~W.}\ \bibnamefont
  {Knight}},\ }\href@noop {} {\bibfield  {journal} {\bibinfo  {journal} {Neural
  Comput.}\ }\textbf {\bibinfo {volume} {12}},\ \bibinfo {pages} {473}
  (\bibinfo {year} {2000})}\BibitemShut {NoStop}%
\bibitem [{\citenamefont {Shiino}\ and\ \citenamefont
  {Fukai}(1993)}]{Shiino1993}%
  \BibitemOpen
  \bibfield  {author} {\bibinfo {author} {\bibfnamefont {M.}~\bibnamefont
  {Shiino}}\ and\ \bibinfo {author} {\bibfnamefont {T.}~\bibnamefont {Fukai}},\
  }\href {https://doi.org/10.1103/PhysRevE.48.867} {\bibfield  {journal}
  {\bibinfo  {journal} {Phys.\ Rev.\ E}\ }\textbf {\bibinfo {volume} {48}},\
  \bibinfo {pages} {867} (\bibinfo {year} {1993})}\BibitemShut {NoStop}%
\bibitem [{\citenamefont {Cohen}\ and\ \citenamefont
  {Grossberg}(1983)}]{Cohen1983}%
  \BibitemOpen
  \bibfield  {author} {\bibinfo {author} {\bibfnamefont {M.~A.}\ \bibnamefont
  {Cohen}}\ and\ \bibinfo {author} {\bibfnamefont {S.}~\bibnamefont
  {Grossberg}},\ }\href {https://doi.org/10.1109/TSMC.1983.6313075} {\bibfield
  {journal} {\bibinfo  {journal} {IEEE Trans.\ Syst.\ Man Cybern.}\ }\textbf
  {\bibinfo {volume} {SMC-13}},\ \bibinfo {pages} {815} (\bibinfo {year}
  {1983})}\BibitemShut {NoStop}%
\bibitem [{\citenamefont {Hopfield}(1984)}]{Hopfield1984}%
  \BibitemOpen
  \bibfield  {author} {\bibinfo {author} {\bibfnamefont {J.~J.}\ \bibnamefont
  {Hopfield}},\ }\href {https://doi.org/10.1073/pnas.81.10.3088} {\bibfield
  {journal} {\bibinfo  {journal} {Proc.\ Natl.\ Acad.\ Sci.\ USA}\ }\textbf
  {\bibinfo {volume} {81}},\ \bibinfo {pages} {3088} (\bibinfo {year}
  {1984})}\BibitemShut {NoStop}%
\bibitem [{\citenamefont {K{\"{u}}hn}\ \emph {et~al.}(1991)\citenamefont
  {K{\"{u}}hn}, \citenamefont {B{\"{o}}s},\ and\ \citenamefont {van
  Hemmen}}]{Kuhn1991}%
  \BibitemOpen
  \bibfield  {author} {\bibinfo {author} {\bibfnamefont {R.}~\bibnamefont
  {K{\"{u}}hn}}, \bibinfo {author} {\bibfnamefont {S.}~\bibnamefont
  {B{\"{o}}s}},\ and\ \bibinfo {author} {\bibfnamefont {J.~L.}\ \bibnamefont
  {van Hemmen}},\ }\href {https://doi.org/10.1103/physreva.43.2084} {\bibfield
  {journal} {\bibinfo  {journal} {Phys.\ Rev.\ A}\ }\textbf {\bibinfo {volume}
  {43}},\ \bibinfo {pages} {2084} (\bibinfo {year} {1991})}\BibitemShut
  {NoStop}%
\bibitem [{\citenamefont {Amit}\ \emph
  {et~al.}(1985{\natexlab{b}})\citenamefont {Amit}, \citenamefont {Gutfreund},\
  and\ \citenamefont {Sompolinsky}}]{Amit1985b}%
  \BibitemOpen
  \bibfield  {author} {\bibinfo {author} {\bibfnamefont {D.~J.}\ \bibnamefont
  {Amit}}, \bibinfo {author} {\bibfnamefont {H.}~\bibnamefont {Gutfreund}},\
  and\ \bibinfo {author} {\bibfnamefont {H.}~\bibnamefont {Sompolinsky}},\
  }\href {https://doi.org/10.1103/PhysRevA.32.1007} {\bibfield  {journal}
  {\bibinfo  {journal} {Phys.\ Rev.\ A}\ }\textbf {\bibinfo {volume} {32}},\
  \bibinfo {pages} {1007} (\bibinfo {year} {1985}{\natexlab{b}})}\BibitemShut
  {NoStop}%
\bibitem [{\citenamefont {van Hemmen}\ \emph {et~al.}(1986)\citenamefont {van
  Hemmen}, \citenamefont {Grensing}, \citenamefont {Huber},\ and\ \citenamefont
  {K{\"{u}}hn}}]{VanHemmen1986}%
  \BibitemOpen
  \bibfield  {author} {\bibinfo {author} {\bibfnamefont {J.~L.}\ \bibnamefont
  {van Hemmen}}, \bibinfo {author} {\bibfnamefont {D.}~\bibnamefont
  {Grensing}}, \bibinfo {author} {\bibfnamefont {A.}~\bibnamefont {Huber}},\
  and\ \bibinfo {author} {\bibfnamefont {R.}~\bibnamefont {K{\"{u}}hn}},\
  }\href {https://doi.org/10.1007/BF01308399} {\bibfield  {journal} {\bibinfo
  {journal} {Zeitschrift f{\"{u}}r Physik B Condensed Matter}\ }\textbf
  {\bibinfo {volume} {65}},\ \bibinfo {pages} {53} (\bibinfo {year}
  {1986})}\BibitemShut {NoStop}%
\bibitem [{\citenamefont {Grensing}\ and\ \citenamefont
  {K{\"{u}}hn}(1987)}]{Grensing1987}%
  \BibitemOpen
  \bibfield  {author} {\bibinfo {author} {\bibfnamefont {D.}~\bibnamefont
  {Grensing}}\ and\ \bibinfo {author} {\bibfnamefont {R.}~\bibnamefont
  {K{\"{u}}hn}},\ }\href {https://doi.org/10.1051/jphys:01987004805071300}
  {\bibfield  {journal} {\bibinfo  {journal} {J. Phys. (Paris)}\ }\textbf
  {\bibinfo {volume} {48}},\ \bibinfo {pages} {713} (\bibinfo {year}
  {1987})}\BibitemShut {NoStop}%
\bibitem [{\citenamefont {Krauth}\ \emph {et~al.}(1988)\citenamefont {Krauth},
  \citenamefont {Mezard},\ and\ \citenamefont {Nadal}}]{Krauth1988}%
  \BibitemOpen
  \bibfield  {author} {\bibinfo {author} {\bibfnamefont {W.}~\bibnamefont
  {Krauth}}, \bibinfo {author} {\bibfnamefont {M.}~\bibnamefont {Mezard}},\
  and\ \bibinfo {author} {\bibfnamefont {J.-P.}\ \bibnamefont {Nadal}},\ }\href
  {https://doi.org/10.5555/56123.56124} {\bibfield  {journal} {\bibinfo
  {journal} {Complex Syst.}\ }\textbf {\bibinfo {volume} {2}},\ \bibinfo
  {pages} {387} (\bibinfo {year} {1988})}\BibitemShut {NoStop}%
\bibitem [{\citenamefont {Folli}\ \emph {et~al.}(2017)\citenamefont {Folli},
  \citenamefont {Leonetti},\ and\ \citenamefont {Ruocco}}]{Folli2017}%
  \BibitemOpen
  \bibfield  {author} {\bibinfo {author} {\bibfnamefont {V.}~\bibnamefont
  {Folli}}, \bibinfo {author} {\bibfnamefont {M.}~\bibnamefont {Leonetti}},\
  and\ \bibinfo {author} {\bibfnamefont {G.}~\bibnamefont {Ruocco}},\
  }\href@noop {} {\bibfield  {journal} {\bibinfo  {journal} {Front. Comput.
  Neurosci.}\ }\textbf {\bibinfo {volume} {10}},\ \bibinfo {pages} {144}
  (\bibinfo {year} {2017})}\BibitemShut {NoStop}%
\bibitem [{\citenamefont {Edwards}\ and\ \citenamefont
  {Anderson}(1975)}]{Edwards1975}%
  \BibitemOpen
  \bibfield  {author} {\bibinfo {author} {\bibfnamefont {S.~F.}\ \bibnamefont
  {Edwards}}\ and\ \bibinfo {author} {\bibfnamefont {P.~W.}\ \bibnamefont
  {Anderson}},\ }\href {https://doi.org/10.1088/0305-4608/5/5/017} {\bibfield
  {journal} {\bibinfo  {journal} {J. Phys. F}\ }\textbf {\bibinfo {volume}
  {5}},\ \bibinfo {pages} {965} (\bibinfo {year} {1975})}\BibitemShut {NoStop}%
\bibitem [{\citenamefont {Akemann}\ \emph {et~al.}(2015)\citenamefont
  {Akemann}, \citenamefont {Baik},\ and\ \citenamefont {{Di
  Francesco}}}]{Akemann2015}%
  \BibitemOpen
  \bibfield  {author} {\bibinfo {author} {\bibfnamefont {G.}~\bibnamefont
  {Akemann}}, \bibinfo {author} {\bibfnamefont {J.}~\bibnamefont {Baik}},\ and\
  \bibinfo {author} {\bibfnamefont {P.}~\bibnamefont {{Di Francesco}}},\ }\href
  {https://doi.org/10.1093/oxfordhb/9780198744191.001.0001} {\emph {\bibinfo
  {title} {The Oxford Handbook of Random Matrix Theory}}}\ (\bibinfo
  {publisher} {Oxford University Press},\ \bibinfo {year} {2015})\ p.\ \bibinfo
  {pages} {919}\BibitemShut {NoStop}%
\bibitem [{\citenamefont {Gardner}(1986)}]{Gardner1986}%
  \BibitemOpen
  \bibfield  {author} {\bibinfo {author} {\bibfnamefont {E.}~\bibnamefont
  {Gardner}},\ }\bibfield  {journal} {\bibinfo  {journal} {J. Phys. A}\
  }\textbf {\bibinfo {volume} {19}},\ \href
  {https://doi.org/10.1088/0305-4470/19/16/017} {10.1088/0305-4470/19/16/017}
  (\bibinfo {year} {1986})\BibitemShut {NoStop}%
\bibitem [{\citenamefont {Treves}\ and\ \citenamefont
  {Amit}(1988)}]{Treves1988}%
  \BibitemOpen
  \bibfield  {author} {\bibinfo {author} {\bibfnamefont {A.}~\bibnamefont
  {Treves}}\ and\ \bibinfo {author} {\bibfnamefont {D.~J.}\ \bibnamefont
  {Amit}},\ }\href {https://doi.org/10.1088/0305-4470/21/14/016} {\bibfield
  {journal} {\bibinfo  {journal} {J. Phys. A}\ }\textbf {\bibinfo {volume}
  {21}},\ \bibinfo {pages} {3155} (\bibinfo {year} {1988})}\BibitemShut
  {NoStop}%
\bibitem [{\citenamefont {Singh}(2001)}]{Singh2001}%
  \BibitemOpen
  \bibfield  {author} {\bibinfo {author} {\bibfnamefont {M.~P.}\ \bibnamefont
  {Singh}},\ }\href {https://doi.org/10.1103/PhysRevE.64.051912} {\bibfield
  {journal} {\bibinfo  {journal} {Phys.\ Rev.\ E}\ }\textbf {\bibinfo {volume}
  {64}},\ \bibinfo {pages} {9} (\bibinfo {year} {2001})}\BibitemShut {NoStop}%
\bibitem [{\citenamefont {Coolen}\ and\ \citenamefont
  {Sherrington}(1993)}]{Coolen1993}%
  \BibitemOpen
  \bibfield  {author} {\bibinfo {author} {\bibfnamefont {A.~C.~C.}\
  \bibnamefont {Coolen}}\ and\ \bibinfo {author} {\bibfnamefont
  {D.}~\bibnamefont {Sherrington}},\ }\href
  {https://doi.org/10.1103/PhysRevLett.71.3886} {\bibfield  {journal} {\bibinfo
   {journal} {Phys.\ Rev.\ Lett.}\ }\textbf {\bibinfo {volume} {71}},\ \bibinfo
  {pages} {3886} (\bibinfo {year} {1993})}\BibitemShut {NoStop}%
\bibitem [{\citenamefont {Coolen}\ and\ \citenamefont
  {Sherrington}(1994)}]{Coolen1994}%
  \BibitemOpen
  \bibfield  {author} {\bibinfo {author} {\bibfnamefont {A.~C.~C.}\
  \bibnamefont {Coolen}}\ and\ \bibinfo {author} {\bibfnamefont
  {D.}~\bibnamefont {Sherrington}},\ }\href
  {https://doi.org/10.1103/PhysRevE.49.1921} {\bibfield  {journal} {\bibinfo
  {journal} {Phys.\ Rev.\ E}\ }\textbf {\bibinfo {volume} {49}},\ \bibinfo
  {pages} {1921} (\bibinfo {year} {1994})}\BibitemShut {NoStop}%
\bibitem [{\citenamefont {Fontanari}\ and\ \citenamefont
  {K{\"{o}}berle}(1988)}]{Fontanari1988}%
  \BibitemOpen
  \bibfield  {author} {\bibinfo {author} {\bibfnamefont {J.}~\bibnamefont
  {Fontanari}}\ and\ \bibinfo {author} {\bibfnamefont {R.}~\bibnamefont
  {K{\"{o}}berle}},\ }\href {https://doi.org/10.1051/jphys:0198800490101300}
  {\bibfield  {journal} {\bibinfo  {journal} {J.\ Physique}\ }\textbf {\bibinfo
  {volume} {49}},\ \bibinfo {pages} {13} (\bibinfo {year} {1988})}\BibitemShut
  {NoStop}%
\bibitem [{\citenamefont {Fachechi}\ \emph {et~al.}(2019)\citenamefont
  {Fachechi}, \citenamefont {Agliari},\ and\ \citenamefont
  {Barra}}]{Fachechi2019}%
  \BibitemOpen
  \bibfield  {author} {\bibinfo {author} {\bibfnamefont {A.}~\bibnamefont
  {Fachechi}}, \bibinfo {author} {\bibfnamefont {E.}~\bibnamefont {Agliari}},\
  and\ \bibinfo {author} {\bibfnamefont {A.}~\bibnamefont {Barra}},\ }\href
  {https://doi.org/10.1016/j.neunet.2019.01.006} {\bibfield  {journal}
  {\bibinfo  {journal} {Neural Netw.}\ }\textbf {\bibinfo {volume} {112}},\
  \bibinfo {pages} {24} (\bibinfo {year} {2019})}\BibitemShut {NoStop}%
\bibitem [{\citenamefont {Agliari}\ \emph {et~al.}(2024)\citenamefont
  {Agliari}, \citenamefont {Alemanno}, \citenamefont {Aquaro},\ and\
  \citenamefont {Fachechi}}]{Agliari2024}%
  \BibitemOpen
  \bibfield  {author} {\bibinfo {author} {\bibfnamefont {E.}~\bibnamefont
  {Agliari}}, \bibinfo {author} {\bibfnamefont {F.}~\bibnamefont {Alemanno}},
  \bibinfo {author} {\bibfnamefont {M.}~\bibnamefont {Aquaro}},\ and\ \bibinfo
  {author} {\bibfnamefont {A.}~\bibnamefont {Fachechi}},\ }\href
  {https://doi.org/10.1016/j.neunet.2024.106389} {\bibfield  {journal}
  {\bibinfo  {journal} {Neural Netw.}\ }\textbf {\bibinfo {volume} {177}},\
  \bibinfo {pages} {106389} (\bibinfo {year} {2024})}\BibitemShut {NoStop}%
\bibitem [{\citenamefont {Serricchio}\ \emph {et~al.}(2025)\citenamefont
  {Serricchio}, \citenamefont {Bocchi}, \citenamefont {Chilin}, \citenamefont
  {Marino}, \citenamefont {Negri}, \citenamefont {Cammarota},\ and\
  \citenamefont {Ricci-Tersenghi}}]{Serricchio2025}%
  \BibitemOpen
  \bibfield  {author} {\bibinfo {author} {\bibfnamefont {L.}~\bibnamefont
  {Serricchio}}, \bibinfo {author} {\bibfnamefont {D.}~\bibnamefont {Bocchi}},
  \bibinfo {author} {\bibfnamefont {C.}~\bibnamefont {Chilin}}, \bibinfo
  {author} {\bibfnamefont {R.}~\bibnamefont {Marino}}, \bibinfo {author}
  {\bibfnamefont {M.}~\bibnamefont {Negri}}, \bibinfo {author} {\bibfnamefont
  {C.}~\bibnamefont {Cammarota}},\ and\ \bibinfo {author} {\bibfnamefont
  {F.}~\bibnamefont {Ricci-Tersenghi}},\ }\href
  {https://doi.org/10.1016/j.neunet.2025.107216} {\bibfield  {journal}
  {\bibinfo  {journal} {Neural Netw.}\ }\textbf {\bibinfo {volume} {186}},\
  \bibinfo {pages} {107216} (\bibinfo {year} {2025})}\BibitemShut {NoStop}%
\bibitem [{\citenamefont {Coolen}(2001)}]{Coolen2001b}%
  \BibitemOpen
  \bibfield  {author} {\bibinfo {author} {\bibfnamefont {A.~C.~C.}\
  \bibnamefont {Coolen}},\ }in\ \href@noop {} {\emph {\bibinfo {booktitle}
  {Handbook of Biological Physics}}},\ Vol.~\bibinfo {volume} {4},\ \bibinfo
  {editor} {edited by\ \bibinfo {editor} {\bibfnamefont {F.}~\bibnamefont
  {Moss}}\ and\ \bibinfo {editor} {\bibfnamefont {S.}~\bibnamefont {Gielen}}}\
  (\bibinfo  {publisher} {Elsevier Science B. V.},\ \bibinfo {year} {2001})\
  pp.\ \bibinfo {pages} {597--662}\BibitemShut {NoStop}%
\end{thebibliography}%

\end{document}